\documentclass[11pt]{article}

\usepackage{graphicx}
\usepackage{cite}
\usepackage{amsmath}
\usepackage{url}
\usepackage{verbatim}
\usepackage{caption} 

\renewcommand{\paragraph}[1]{\smallskip\noindent {\bf #1}~}

\begin{document}

\markboth{}{}

\title{Efficient Hybrid Inline and Out-of-line Deduplication for Backup Storage}

\author{Yan-Kit LI, Min Xu, Chun-Ho Ng, and Patrick P. C. Lee\\
The Chinese University of Hong Kong}






\maketitle

\begin{abstract}
Backup storage systems often remove redundancy across backups via inline
deduplication, which works by referring duplicate chunks of the latest backup
to those of existing backups. However, inline deduplication degrades restore
performance of the latest backup due to fragmentation, and complicates
deletion of expired backups due to the sharing of data chunks.  While
out-of-line deduplication addresses the problems by forward-pointing existing
duplicate chunks to those of the latest backup, it introduces additional I/Os
of writing and removing duplicate chunks. 

We design and implement {\em RevDedup}, an efficient hybrid inline and
out-of-line deduplication system for backup storage. It applies coarse-grained
inline deduplication to remove duplicates of the latest backup, and then
fine-grained out-of-line reverse deduplication to remove duplicates from
older backups.  Our reverse deduplication design limits the I/O overhead and
prepares for efficient deletion of expired backups.  Through extensive testbed
experiments using synthetic and real-world datasets, we show that RevDedup can
bring high performance to the backup, restore, and deletion operations, while
maintaining high storage efficiency comparable to conventional inline
deduplication. 
\end{abstract}

\section{Introduction} 
\label{sec:intro} 


Deduplication is an established technique for eliminating data redundancy in
backup storage.  It treats data as a stream of fixed-size or variable-size
{\em chunks}, each of which is identified by a {\em fingerprint} computed by a
cryptographic hash (e.g., MD5, SHA-1) of its content.  Two chunks are said to
be identical if their fingerprints are the same, while fingerprint collisions
of two different chunks are very unlikely \cite{black06}.  Instead of storing
multiple identical chunks, deduplication stores only one unique copy of a
chunk and refers any duplicate copies to the unique copy using smaller-size
references.  Since backups have high redundant content, it is reported that
deduplication can help backup systems achieve effective storage saving by
20$\times$ \cite{andrews13}. 

\subsection{Inline vs. Out-of-line Deduplication}

Deduplication can be realized {\em inline}, which removes duplicate chunks on
the write path, or {\em out-of-line}, which first stores all data and later
removes duplicates in the background.  Today's production backup systems
\cite{zhu08,wallace12,lillibridge13}, which mainly build on disk-based
backends, often implement inline deduplication with average chunk size
4$\sim$8KB. 
However, inline deduplication poses several fundamental challenges to the
basic operations of backup systems, including backup, restore, and deletion: 
\begin{itemize} \itemsep=3pt \parskip=3pt
\item
{\em Backup:}  While inline deduplication avoids writing duplicates, its
backup performance can be degraded by extensive metadata operations for chunk
indexing, including fingerprint computations and index updates. The amount of
metadata increases proportionally with the number of chunks stored.  Thus,
keeping all fingerprints and other metadata in main memory is infeasible.
Instead, some indexing metadata must be kept on disk, but this incurs disk
accesses for metadata lookups and degrades backup performance.
\item
{\em Restore:}  Inline deduplication introduces {\em fragmentation}
\cite{rhea08,kaczmarczyk12,nam12,lillibridge13}, as backups now refer to
existing data copies scattered in prior backups.  This incurs significant
disk seeks when restoring recent backups, and the restore performance
degrades. 
Fragmentation becomes worse for newer backups, whose data is scattered
across more prior backups.  The gradual degradation is undesirable since the
new backups are more likely to be restored during disaster recovery.  A lower
restore throughput of the latest backup implies a longer system downtime. 
\item
{\em Deletion:}  With inline deduplication, expired backups cannot be directly
deleted as they may be shared by newer, non-expired backups.  Deletion is
often handled via a mark-and-sweep approach: in the mark phase, all chunks are
scanned and any unreferenced chunks are marked for removal; in the sweep
phase, all marked chunks are freed from disk in the background.   However, the
mark phase needs to search for unreferenced chunks across disk and incurs
significant I/Os.  
\end{itemize}

Extensive studies address the above challenges of inline deduplication (see
Section~\ref{sec:related}).  However, it remains an open issue of how to
address the challenges {\em simulataneously} so as to enable
deduplication-enabled backup systems to achieve high performance in backup,
restore, and deletion operations. 

Out-of-line deduplication addresses some aforementioned issues of inline
deduplication.  For example, it can reduce the disk I/O overhead of index
lookups on the write path.  It also mitigates fragmentation and preserves
restore performance of the new backups by referring duplicate chunks of old
backups to the chunks of new backups \cite{kaczmarczyk12}.  This
forward-pointing approach also facilitates the deletion of old backups, since
their chunks are no longer shared by new backups.  However, out-of-line
deduplication incurs extra I/Os of writing and removing redundant data, and
hence gives poorer backup performance than inline
deduplication.  For example, writing duplicates can slow down the backup
performance by around 3$\times$ compared to inlne deduplication based on the
measurements in a commercial backup system \cite{kaczmarczyk12}.  Also,
out-of-line deduplication needs extra storage space to keep redundant data
before the redundant data is removed. 


\subsection{Contributions}

Our position is that both inline deduplication and out-of-line deduplication
complement each other if carefully used.  We propose {\em RevDedup}, an
efficient hybrid inline and out-of-line deduplication system for backup
storage.  Our work extends our prior work \cite{ng13}
to aim for high performance in backup, restore, and deletion operations, while
preserving storage efficiency as in conventional inline deduplication.
RevDedup first applies coarse-grained inline deduplication at the granularity
of large-size units, and further applies fine-grained out-of-line
deduplication on small-size units to improve storage efficiency.  Our
out-of-line deduplication step, called {\em reverse deduplication}, shifts
fragmentation to older backups by referring their duplicates to newer
backups.  To limit the I/O overhead of reverse deduplication, we compare only 
two consecutive backup versions derived from the same client, and we argue
that it still effectively removes duplicates.  Also, during reverse
deduplication, we repackage backup data to facilitate subsequent deletion of
expired backups.

We implement a RevDedup prototype and conduct extensive testbed experiments
using synthetic and real-world workloads.  We show that RevDedup maintains
comparable storage efficiency to conventional inline deduplication, achieves
high backup and restore throughput for recent backups (e.g., on the order of
GB/s), and supports fast deletion for expired backups.  To our knowledge, very
few deduplication studies in the literature evaluate the actual I/O performance
through prototype implementation. 

The rest of the paper proceeds as follows.  In Section~\ref{sec:design} and
Section~\ref{sec:implementation}, we present the design and implementation
details of RevDedup, respectively.  In Section~\ref{sec:evaluation}, we report
testbed experimental results.  We review related work in
Section~\ref{sec:related}, and finally conclude the paper in
Section~\ref{sec:conclusion}. 

\section{RevDedup Design} 
\label{sec:design}

RevDedup combines inline and out-of-line deduplication and is designed for
backup storage.  It aims for the following design goals: 
\begin{itemize}
\item
Comparable storage efficiency to conventional inline deduplication
approaches;  
\item
High backup throughput for new backups; 
\item
High restore throughput for new backups; and
\item 
Low deletion overhead for expired backups.
\end{itemize}

\subsection{Backup Basics}

{\em Backups} are copies of primary data snapshotted from client systems or
applications, and can be represented in the form of {\tt tar} files or VM disk
images (e.g., {\tt qcow2}, {\tt vmdk}, etc.).  They are regularly created by a
backup system, either as daily incremental backups or weekly full backups.
Backup data is organized into {\em containers} as the units of storage and
read/write requests, such that each container is of size on the order of
megabytes.  Today's backup solutions mainly build on disk-based storage, which
achieves better I/O performance than traditional tape-based storage.

We define a {\em series} as the sequence of backups snapshotted from the same
client at different times.  Each backup series has a retention period
\cite{wallace12}, which defines how long a backup is kept in storage.  We
define a {\em retention window} that specifies a set of recent backups that
need to be kept in storage.  The retention window slides over time to cover
the latest backup, while the earliest backup stored in the system expires.
The backup system later deletes the expired backups and reclaims storage
space.   Note that the retention window length may vary across different
backup series. 

Since backups share high redundancy, this work focuses on using deduplication
to remove redundancy and achieve high storage efficiency.  We can further
improve storage efficiency through local compression (e.g., Ziv-Lempel
\cite{ziv77}), yet we do not consider the effect of compression in this work. 

\subsection{RevDedup Overview}
\label{subsec:goal}

RevDedup performs deduplication in two phases.  It first applies inline
deduplication by dividing backup data into large-size units (e.g., on the
order of megabytes) called {\em segments}, and removes duplicate segments of
new backups on the write path.  It packs the unique segments into containers
and stores the containers on disk.  Deduplication on large-size segments
reduces both fragmentation and indexing overheads \cite{kruus10}.  See
Section~\ref{subsec:seg_dedup} for details. 

RevDedup then reads the containers and applies out-of-line deduplication to
small-size data units (e.g., on the order of kilobytes) called {\em chunks}.
It further removes duplicate chunks of older backups and refers them to the
identical chunks in newer backups.  We call this {\em reverse deduplication},
which shifts fragmentation to older backups and hence maintains high restore
throughput of newer backups.   See Section~\ref{subsec:chunk_dedup} for
details. 

After reverse deduplication, RevDedup repackages segments into separate
containers to facilitate later deletions of expired backups.  See
Section~\ref{subsec:delete} for details. 


\begin{figure}[t]
\centering
\includegraphics[width=0.8\linewidth]{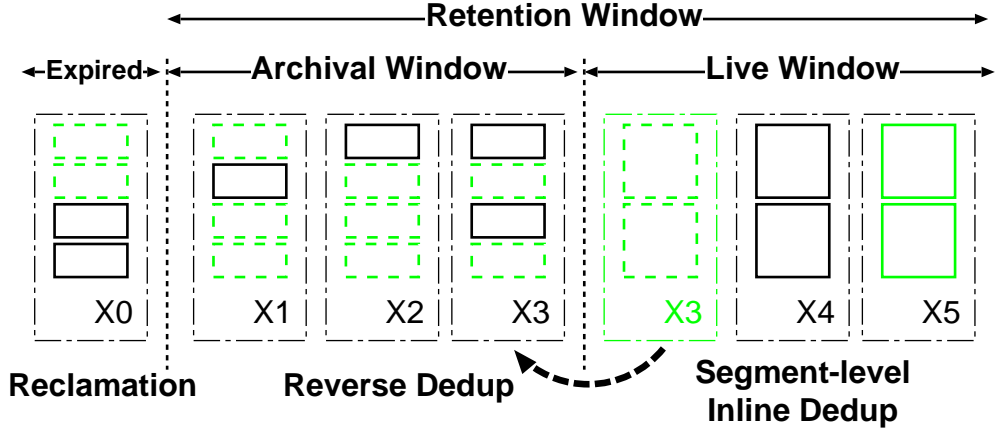}
\caption{Overview of RevDedup operations.}
\label{fig:overview}
\end{figure}

We first describe how RevDedup prepares for the two-phase deduplication before 
explaining the design details. 

\subsubsection{Live and Archival Backups}

RevDedup divides the retention window of a backup series into two sub-windows:
{\em live window} and {\em archival window}.  Backups in the live window
(called {\em live backups}) are those recently written and are more likely to
be restored, while those in the archival window (called 
{\em archival backups}) serve for the archival purpose only and are rarely
restored.  RevDedup applies inline deduplication to the latest backup, which
is first stored in the live window.  The retention window then slides to cover
the latest backup.  The oldest live backup will move to the archival window,
and RevDedup applies reverse deduplication to that backup out-of-line (e.g.,
when the storage system has light load).  

Figure~\ref{fig:overview} illustrates the lifecycles of
six backups of the same series created in the following order: X0, X1, X2, X3,
X4, and X5.  Suppose that the retention window is set to five backups, the
live window is set to two backups, and the archival window is set to three
backups. When X5 is added,  X0 expires and can be deleted to reclaim disk
space.  Also, the segments of X5 can be deduplicated with those of existing
live backups (i.e., X4 in this example).  Also, X3 moves to the
archival window.  We can perform reverse deduplication and remove duplicate
chunks from X3. 

\subsubsection{Chunking}

{\em Chunking} is the process of dividing a data stream into fixed-size or
variable-size deduplication units (i.e., segments or chunks).  Our discussion
assumes variable-size chunking.  Here, we consider the chunking approach based
on Rabin Fingerprinting \cite{rabin81}, whose idea is to compute a rolling
hash of a sliding chunking window over the data stream and then identify
boundaries whose lower-order bits of the rolling hash match a target pattern.
An important property of Rabin Fingerprinting is that the new rolling hash
value can be efficiently computed using the last rolling hash value. 

RevDedup identifies both segment and chunk boundaries using the same target
pattern.  We define two bit lengths $m$ and $n$, where $m > n$, which
correspond to the average segment and chunk sizes, respectively.  When the
chunking window slides, we first check whether the lowest $n$ bits of the
rolling hash match the target pattern. If yes, the window endpoint is a chunk
boundary, and we further check whether the lowest $m$ bits of the rolling hash
match the same target pattern; if yes, the window endpoint is also a segment
boundary.  Clearly, a segment boundary must also be a chunk boundary.  The
chunking process can be done in a single pass of the data stream, and hence
preserves the chunking performance of Rabin Fingerprinting.

In our discussion, the segment or chunk size configured in variable-size
chunking actually refers to an average size.  We also assume that the minimum
and maximum segment or chunk sizes are half and twice the average size,
respectively.  


\subsection{Segment-level Inline Deduplication} 
\label{subsec:seg_dedup} 

RevDedup performs segment-level inline deduplication to the storage pool. 
As in conventional inline deduplication, RevDedup performs deduplication 
{\em globally} in different levels: within the same backup, across different
backups of the same series, and across different series of backups.  The main
difference is on the deduplication unit size: RevDedup uses large-size units
(called segments) on the order of megabytes (e.g., 4$\sim$8MB), while
conventional inline deduplication uses small-size units on the order of
kilobytes (e.g., 4KB \cite{guo11} or 8KB \cite{zhu08}). 

Choosing large deduplication units (segments) has two key benefits. First, it
mitigates fragmentation \cite{srinivasan12}.  Since we put the entire
segment in a container, we reduce the number of containers that need to be
accessed with a large segment size.  In addition, it keeps a small
deduplication index (i.e., the data structure for holding the segment
fingerprints and their locations), and hence mitigates the indexing overhead
\cite{kruus10}.  For example, suppose that we store 1PB of data, the segment
size is 4MB, and the index entry size is 32~bytes.  Then the index size is 
8GB only, as opposed to 8TB when the deduplication unit size is 4KB. 
	
Segment-level inline deduplication still achieves reasonably high
deduplication efficiency, as changes of backups are likely aggregated in
relatively few small regions, while several extended regions remain the same
\cite{kruus10}. 
Nevertheless, using large deduplication units cannot maintain the same level
of deduplication efficiency as do conventional fine-grained deduplication
approaches (see our evaluations in Section~\ref{sec:evaluation}).

To keep track of the deduplication metadata for all stored segments, our
current RevDedup design maintains an in-memory deduplication index.  We can
keep an on-disk index instead to reduce memory usage, and exploit compact data
structures and workload characteristics to reduce on-disk index lookups
\cite{zhu08}.  Another option is to keep the index on solid-state drives
\cite{debnath10,meister10}.  The issues of reducing memory usage of indexing
are posed as future work. 

RevDedup packs the unique segments into a fixed-size container. To handle
variable-size segments, we initialize a new container with a new segment (even
the segment size is larger than the container size).  We then keep adding new
segments to the container if it is not full.  If adding a segment exceeds the
container size, we seal and store the container, and create a new container
for the segment being added. 




\subsection{Reverse Deduplication} 
\label{subsec:chunk_dedup}

After segment-level inline deduplication, RevDedup divides each segment into
smaller-size chunks, each of which has size on the order of kilobytes, and
applies chunk-level out-of-line deduplication to further improve storage
efficiency.  To mitigate fragmentation of the newer backups that are more
likely to be restored,  RevDedup performs reverse deduplication, i.e.,
removing duplicate chunks in older backups and referring them to
identical chunks in newer backups. 

However, we must address several issues of out-of-line deduplication. First,
there are extra I/Os of identifying and removing duplicate chunks on disk. To
reduce the I/O overhead, we limit the deduplication operation to two
consecutive backups of the same series.  Second, we can remove duplicate chunks
only when their associated segments are no longer shared by other backups.
We use two-level reference management to keep track of how segments and chunks
are referenced and decide if a chunk can be safely removed.  Third, we must
support efficient removal of duplicate chunks.  We argue that we only check 
the segments that are not shared by any live backups for chunk removal. 
In the following, we describe how we address the issues altogether.

\subsubsection{Deduplication Operation}
\label{subsubsec:reverse}

Our reverse deduplication works on two consecutive backups of the same series.
When a live backup (call it X0) moves from the live window to the archival
window, RevDedup loads the metadata of the following (live) backup of the same
series (call it X1).  It then removes duplicate chunks from X0 and refers them
to those in X1.  

The deduplication operation follows two principles and we provide
justifications.  First, we limit reverse deduplication to the {\em backups of
the same series}.  Due to repeated backups of the same client system,
inter-version duplicates of the same series are common \cite{kaczmarczyk12}.
Also, changes of a backup tend to appear in small regions \cite{kruus10}.
Thus, we can potentially remove additional inter-version duplicates around the
small change regions in a more fine-grained way \cite{kruus10}.  Second, 
reverse deduplication is applied to {\em consecutive backups}.  Our assumption
is that most duplicate chunks appear among consecutive backups. 
Our current design focuses on only two consecutive backups, yet we can
compare more backups to trade deduplication performance for storage efficiency. 

\begin{figure}[t]
\centering
\includegraphics[width=0.8\linewidth]{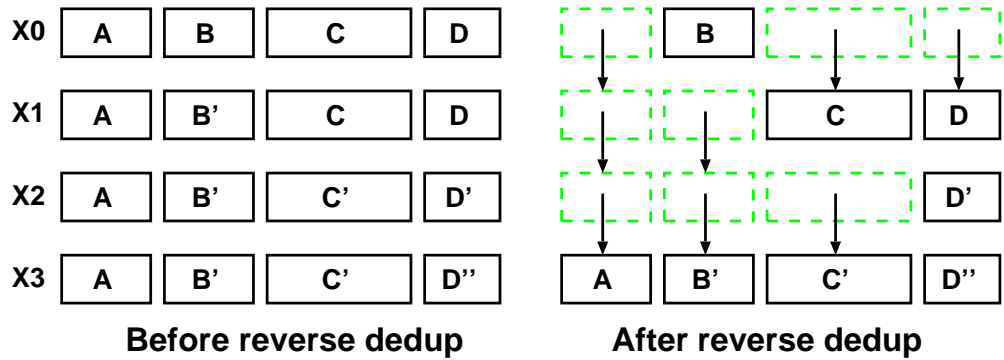}
\caption{Example of reverse deduplication for four backups of the same series
created in the order: X0, X1, X2, and X3.}
\label{fig:revdedup}
\end{figure}

Each backup keeps a list of references to all chunks.  Each chunk reference is
one of the two types: either (1) a {\em direct reference}, which points to a
physical chunk, or (2) an {\em indirect reference}, which points to a
reference of the following backup of the same series.  Since the following
backup may be further deduplicated with its own following backup, accessing a
chunk may follow a chain of indirect references.  Figure~\ref{fig:revdedup}
shows an example of reverse deduplication for four backups created in the
order X0, X1, X2, and X3.  We see that X0 (the oldest backup) may access a
chunk of X1 through an indirect reference, or a chunk of X2 or X3 through a
chain of indirect references.  Note that the latest backup must have direct
references only. 

To perform reverse deduplication between the old backup X0 and the following
backup X1, RevDedup loads the chunk fingerprints of X1 from the metadata
store (see Section~\ref{sec:implementation}) and builds an in-memory index on the
fly.  It then loads the chunk fingerprints of X0 and checks if they match any
chunk fingerprints of X1 in the index.  We quantify the worst-case memory
usage as follows.  Suppose that the raw size of a backup is 20GB, the chunk
size is 4KB, and each chunk-level index entry size is 32~bytes.  The total
memory usage is up to 20GB$\div$4KB$\times$32~bytes = 160MB.  
Note that the index only temporarily resides in memory and will be discarded
after we finish reverse deduplication. 


\subsubsection{Two-level Reference Management}
\label{subsubsec:twolevel}

After reverse deduplication, we can remove chunks that are not referenced by
any backup from disk to reclaim storage space.  RevDedup uses two-level
reference management to keep track of how segments and chunks are shared. 

RevDedup associates each segment with a reference count,
which indicates the number of references from the live backups. 
Suppose that we now store a segment of a new backup.  If the segment is
unique, its reference count is initialized as one; if it is a duplicate, its
corresponding reference count is incremented by one.  When a live backup moves
to the archival window, all its associated segments have their
reference counts decremented by one.  Reverse deduplication is only applied to
segments that have zero reference counts, meaning that the segments are not
shared by any live backup, and hence their chunks can be removed.  To simplify
our discussion, we call the segments with zero reference counts 
{\em non-shared}, and those with positive reference counts {\em shared}.
We only check the non-shared segments for chunk removal.

If a chunk is found duplicate in the next backup, an indirect reference is
recorded; if a chunk is unique, a direct reference is set.  A chunk can be
safely removed if both conditions hold: (1) its associated segment is
non-shared, and (2) it holds an indirect reference. 

\begin{figure}[t]
\centering
\includegraphics[width=0.8\linewidth]{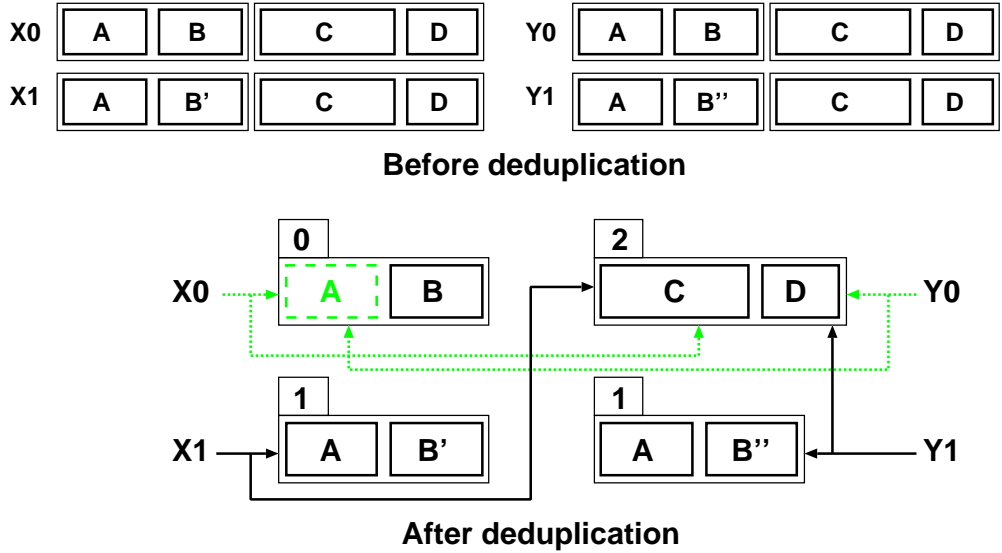}
\caption{Example of reference management in RevDedup for two backup series
\{X0,X1\} and \{Y0,Y1\}, where X0 and Y0 are in the archival window, while X1
and Y1 are in the live window.}
\label{fig:reference}
\end{figure}

Figure~\ref{fig:reference} shows how we manage the segment and chunk
references.  Suppose that we store the backups \{X0, X1\} and \{Y0, Y1\} of
two separate backup series.  Also, we assume that when X1 and Y1 are stored
in the live window, both X0 and Y0 move to the archival window.  Let the
segment size be two chunks.   From the figure, the segment {\sf AB} is no
longer shared by any live backup, so its segment count is zero. 
Also, X0 can refer to chunk {\sf A} in X1, so chunk {\sf A} can be removed
from X0 by reverse deduplication.  Since the segment {\sf CD} is still shared
by X1 and Y1, its reference count is two.  Both segments {\sf AB'} and 
{\sf AB'' } have reference counts equal to one. 

\subsubsection{Chunk Removal}
\label{subsubsec:removal}

RevDedup loads the containers that have non-shared segments.  It compacts all
non-shared segments without the removed chunks in reverse deduplication, and
repackages them into separate containers.  It also rewrites the loaded
containers with the remaining shared segments with positive reference counts
back to disk.  
Separating the non-shared and shared segments into different containers has
two benefits.  First, when we run the chunk removal process next
time, the repackaged containers with non-shared segments are untouched.  This
saves the unnecessary I/Os.  Second, it supports efficient deletion of expired
backups, as described in Section~\ref{subsec:delete}.




\subsection{Deletion of Backups} 
\label{subsec:delete}

RevDedup supports efficient deletion of expired backups using 
{\em container timestamps}.  When it removes chunks from non-shared segments
and repackages them into a new container (see
Section~\ref{subsubsec:removal}), it
associates with the container a timestamp that specifies the creation time of
the corresponding backup.  Any segments whose backups are created at about the
same time can be gathered and packed into the same container, even though the
backups may be derived from different series.  For containers with shared
segments, their timestamps are set to be undefined. 

To delete expired backups, RevDedup examines the well-defined timestamps of
all containers and delete the expired containers.  Such containers must
contain non-shared segments that belong to expired backups, and hence are safe
to be deleted.  We do not need to scan all segments/chunks as in the
traditional mark-and-sweep approach, so the deletion time is significantly
reduced. 



\section{Implementation} 
\label{sec:implementation}

We have implemented a RevDedup prototype in C on Linux.  The prototype mounts
its storage backend on a native file system, which we choose to be Linux Ext4
in this work.  In this section, we describe the components of the prototype,
including metadata and executable modules.  We also present techniques that
further improve the prototype performance. 


\subsection{Metadata}

We maintain deduplication metadata for each of the segments, chunks,
containers, and backup series: (1) the metadata of each segment describes the
segment fingerprint, the fingerprints of all chunks in the segment, the
reference count (for chunk removal in reverse deduplication), and the segment
offset; (2) the metadata of each chunk describes the chunk fingerprint and the
chunk offset; (3) the metadata of each container describes segments in the
container and the timestamp (for reclamation); and (4) the metadata of each
backup series describes which versions are in the live, archival, and
retention windows.  We store each type of the metadata as a log-structured
file with fixed-size entries, each of which is indexed by a unique identifier.
We map the metadata logs into memory using {\tt mmap()}, so the entries are
loaded into memory on demand. 

In addition, we maintain an in-memory deduplication index for segment-level
inline deduplication (see Section~\ref{subsec:seg_dedup}).  We implement the index
as a hash map using the Kyoto Cabinet library \cite{kyotocab}.  Each segment
fingerprint is then mapped to an identifier in the segment metadata log.  In
our prototype, we compute the segment and chunk fingerprints with SHA-1 using
the OpenSSL library \cite{openssl}. 


Each backup is associated with a recipe that contains a list of references for
reconstructing the backup.   For a live backup, the recipe describes the
references to the unique segments; for an archival backup, the recipe holds
both direct and indirect references, which state the offsets of chunks on disk
and the offsets of direct reference entries, respectively. 


\subsection{Executable Modules}
\label{subsec:modules}

We decompose the RevDedup prototype into executable modules that run as
standalone programs.  We can also run them as daemons and connect them via
inter-process communication.  
\begin{itemize}  \itemsep=0pt \parskip=0pt
\item
{\sf chunking:} It chunks a backup file into segments and chunks, and stores
the fingerprints and offsets in a temporary file. 
\item
{\sf inlinededup:}  It performs segment-level inline deduplication on the
backup file, using the temporary file created from {\sf chunking}.  It first
loads the in-memory segment deduplication index from the segment metadata log.
For new unique segments, it adds them into containers, appends metadata to the
segment metadata log, and adds new entries to the deduplication index.  It
also creates the backup recipe holding all the segment references for the
backup. 
\item
{\sf revdedup:}  It takes a backup of a series as the input and performs
reverse deduplication on itself and its following backup of the same series.
It also repacks segments with removed chunks into different containers. 
\item
{\sf restore:} It reconstructs the chunks of a backup given the series and
version numbers as inputs.  It reads the backup recipe, and returns the chunks
by tracing the direct references or chains of indirect references. 
\item
{\sf delete:}  It takes a timestamp as the input and deletes all backups
created earlier than the input timestamp. 
\end{itemize}

\subsection{Further Improvements}
\label{subsec:improvements}

We present techniques that improve the performance of RevDedup during
deployment. 

\paragraph{Multi-threading:} RevDedup exploits multi-threading to parallelize 
operations.  For example, during backup, multiple threads check the segments
for inline deduplication opportunities and write segments into containers; 
during reverse deduplication, multiple threads read containers and check for
chunk removal; during restore, multiple threads read containers and trace
indirect reference chains to reconstruct the segments. 

\paragraph{Prefetching:} RevDedup reads containers during reverse
deduplication and restore.  It uses prefetching to improve read performance.
Specifically, a dedicated prefetch thread calls the POSIX function {\tt
posix\_fadvise(POSIX\_FADV\_WILLNEED)} to notify the kernel to prefetch the
containers into cache and save future disk reads.  While the prefetch thread 
issues the notification and waits for the response from the kernel, other
threads work on metadata processing and data transmission so as to mitigate
the notification overhead.  Note that prefetching is also used by Lillibridge
{\em et al.} \cite{lillibridge13} (known as the forward assembly area) to
improve read performance of deduplication systems.  Our prefetching approach
differs in that it leverages the kernel support. 




\paragraph{Handling of null chunks:}  Some backup workloads such as VM images
may contain a large number of null (or zero-filled) chunks \cite{jin09}. 
RevDedup skips writing null chunks.  When a read request is issued to a null
chunk, the {\sf restore} module returns the null chunk on the fly instead of
reading it from disk.  This improves both backup and restore performance. 

\paragraph{Tunable parameters:} 
RevDedup makes performance trade-offs through configurable parameters,
including the sizes of segments, chunks, and containers, as well as the
lengths of retention, live, archival windows.  For example, a longer live
window implies that more backups are ready to be restored, while consuming
more storage space; larger segments and chunks imply less indexing overhead
and data fragmentation, while reducing deduplication efficiency.  We explore
the performance effects of different parameters in
Section~\ref{sec:evaluation}. 


\section{Experiments}
\label{sec:evaluation}

We conduct testbed experiments on our RevDedup prototype.  We show that
RevDedup achieves high storage efficiency, high backup throughput, high
restore throughput of new backups, and low deletion overhead of expired
backups. 

\subsection{Setup}

\paragraph{Datasets:} 
We evaluate RevDedup using both synthetic and real-world datasets.  For
synthetic datasets, we extend the idea by Lillibridge {\em et al.}
\cite{lillibridge13} to generate configurable workloads for stress-testing
data fragmentation.  
We simulate a backup series by first creating a full backup using a
Ubuntu~12.04 virtual machine (VM) disk image configured with 8GB space.
Initially, the image has 1.1GB of system files.  On each simulated weekday, we
randomly walk through the file system to pick $\alpha\%$ of files and modify
$\beta\%$ of file contents, and further add $\gamma$MB of new files to the file
system.  The parameters $\alpha$, $\beta$, and $\gamma$ are configurable in our
evaluation.  We represent five simulated weekdays as one simulated week.  At
the start of each simulated week, we perform a full backup of the disk image
using the {\tt dd} utility.  We generate 78 full backups to simulate a
duration of 1.5~years.  We configure the parameters to simulate five types of
activities of a single backup series, as listed in Table~\ref{tab:datasets}.
We call the datasets \textit{SG1-5}.  In addition, we also simulate a scenario
with a group of 16 backup series covering 20 weekly full backups each.  We
call the dataset \textit{GP}. 


\begin{table}
\centering
\begin{tabular}{|c||c|c||c|c|c|}
\hline
Traces & \# Series & \# Backups & $\alpha\%$ & $\beta\%$ & $\gamma$MB \\
\hline
\textit{SG1} & 1 & 78 & 2\% & 10\% & 10MB \\
\hline
\textit{SG2} & 1 & 78 & 4\% & 10\% & 10MB\\
\hline
\textit{SG3} & 1 & 78 & 2\% & 20\% & 10MB\\
\hline
\textit{SG4} & 1 & 78 & 2\% & 10\% & 20MB\\
\hline
\textit{SG5} & 1 & 78 & 10\% & 10\% & 10MB\\
\hline
\textit{GP} & 16 & 320 & 2\% & 10\% & 10MB\\
\hline
\end{tabular}
\caption{Details of all synthetic datasets.}
\label{tab:datasets}
\end{table}

We also consider a real-world dataset taken from the snapshots of VM images
used by university students in a programming course.  We prepared a master
image of 7.6GB installed with Ubuntu~10.04 and assigned it to each student to
work on three programming assignments over a 12-week span.  We took weekly
snapshots for the VMs.  For privacy reasons, we only collected cryptographic
hashes on 4KB fixed-size blocks.  For our throughput tests, we reconstruct
disk blocks derived from the hashes via a one-to-one function.  Our evaluation
selects a subset of 80 VMs covering a total of 960 weekly full backups.  The
total size is 7.2TB with 3.3TB of non-zero blocks.  We call the dataset
\textit{VM}.  Note that the dataset only presents a special real-world use
case, and we do not claim its representativeness for general virtual desktop
environments. 

\paragraph{Testbed:} 
We conduct our experiments on a machine with an Intel Xeon E3-1240v2
quad-core, eight-threaded processor, 32GB RAM, and a disk array with eight
ST1000DM003 7200RPM 1TB SATA disks.  By default, we configure a RAID-0 array
as done in prior work \cite{guo11} to maximize the disk array throughput for
high-performance tests, while we also consider RAID-5 and RAID-6 in baseline
performance tests (see Section~\ref{subsec:baseline}).  We fix the RAID chunk
size at 512KB.  The machine runs Ubuntu 12.04.3 with Linux kernel 3.8. 

\paragraph{Default settings:}
We compare RevDedup and conventional inline deduplication. 
For RevDedup, we fix the container size at 32MB, the segment size at 4MB for
inline deduplication, the chunk size at 4KB for reverse deduplication.  We
also assume that the retention window covers all backups. We fix the live
window length to be one backup and the archival window length to be the number
of all remaining backups.  For conventional inline deduplication, we configure
RevDedup to fix the segment size as 4KB and disable reverse deduplication.
The container size is also fixed at 32MB.  We refer to conventional inline
deduplication as \textit{Conv} in the following discussion.  

For the datasets \textit{SG1-5} and \textit{GP}, both RevDedup and
\textit{Conv} use variable-size chunking based on Rabin Fingerprinting
\cite{rabin81}; for the dataset \textit{VM}, we use fixed-size chunking, which
is known to be effective for VM image storage \cite{jin09}.  We examine the
effects of various parameters, including the container size, the segment size,
and the live window length.   

\paragraph{Evaluation methodology:} 
Our experiments focus on examining the I/O performance of RevDedup.  When we
perform throughput and latency measurements, we exclude the overhead of
fingerprint computations, which we assume can be done by backup clients
offline before they store backup data.  We pre-compute all segment and chunk
fingerprints before benchmarking.  In addition, for each write, we call 
{\tt sync()} to force all data to disk.  Before each read, we flush the file
system cache using the command ``{\tt echo 3 > /proc/sys/vm/drop\_caches}''.
By default, we disable prefetching (see Section~\ref{subsec:improvements}) to focus
on the effect of disk accesses on I/O performance. 



\subsection{Baseline Performance}
\label{subsec:baseline}

We measure the baseline performance of RevDedup using unique data (i.e.,
without any duplicates).  We write 8GB of unique data, and then read the same
data from disk.  We also measure the raw throughput of the native file system. 
We obtain averages and standard deviations over 20 runs. 

Table~\ref{tab:raid_benchmark} shows the results.  In RAID-0, RevDedup can
achieve at least 95.9\% and 88.6\% of raw write and read throughputs,
respectively.  We also configure the testbed as RAID-5 and RAID-6.  We observe
throughput drops due to the storage of parities.  Nevertheless, RevDedup
achieves nearly the raw throughput.		


\begin{table}[t]
\centering
\begin{tabular}{|c||c|c|}
\hline
(GB/s) & Raw &  RevDedup \\ 
\hline
W (R0) & 1.060 (0.013) & 1.017 (0.034) \\ 
R (R0) & 1.235 (0.004) & 1.094 (0.004) \\ 
\hline
W (R5) & 0.913 (0.011) & 0.81 (0.013) \\ 
R (R5) & 1.004 (0.020) & 0.85 (0.008) \\ 
\hline
W (R6) & 0.793 (0.016) & 0.734 (0.020) \\ 
R (R6) & 0.935 (0.010) & 0.726 (0.029) \\ 
\hline
\end{tabular}
\caption{Baseline throughput of RevDedup with segment size 4MB and container
size 32MB on unique data under RAID-0, RAID-5, and RAID-6 (values in the
brackets are standard deviations).}
\label{tab:raid_benchmark}
\end{table}

\subsection{Storage Efficiency}
\label{subsec:storage}

We calculate the percentage reduction of storage space with deduplication.
We exclude the metadata overhead and null chunks in our calculation.  We
compare RevDedup and \textit{Conv}.  For RevDedup, we vary the segment sizes.
We also provide a breakdown for segment-level inline deduplication and reverse
deduplication.  

\begin{figure}[t]
\centering
\includegraphics[width=4in]{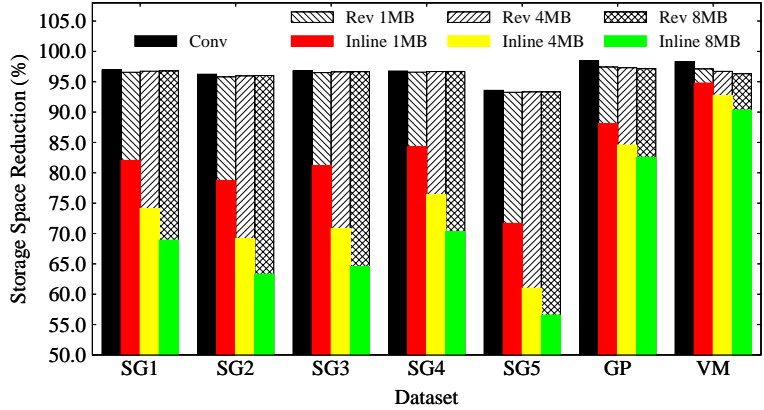}
\caption{Percentage reduction of storage space of RevDedup and \textit{Conv}.}
\label{fig:dedup_ratio}
\end{figure}

Figure~\ref{fig:dedup_ratio} shows the results.  Consider the synthetic
datasets \textit{SG1-5} and \textit{GP}.  In RevDedup, segment-level inline
deduplication itself also reduces storage space, but the saving drastically
drops as the segment size increases.  For example, when the segment size is
8MB, segment-level inline deduplication only gives 56.5$\sim$68.9\% of space
saving for \textit{SG1-5}.  Nevertheless, reverse deduplication increases the
saving to 93.6$\sim$97.0\%, which is comparable to \textit{Conv}.  
	
For the real-world dataset \textit{VM}, RevDedup achieves a saving of
96.3$\sim$97.1\%, which is close to 98.3\% achieved by \textit{Conv}.  In
particular, segment-level inline deduplication saves at least 90\% of space,
since most system files remain unchanged in the VM images.  We emphasize that
the findings are specific to our dataset and may not hold in general. 


\subsection{Throughput}
\label{subsec:throughput}

We evaluate the backup and restore throughput of RevDedup, and compare the
results with \textit{Conv}.  We study the how different segment sizes and
container sizes affect the backup and restore throughput.  We only focus on
the datasets \textit{SG1}, \textit{GP}, and \textit{VM}.  We also study the
overhead of reverse deduplication, the gains of prefetching, and the effect of
live window length, where we focus on the dataset \textit{SG1}.  The results
for \textit{SG1} are plotted every three weeks for clarify of presentation. 
All results are averaged over five runs. 


\subsubsection{Backup}
\label{subsubsec:backup_eval}

\begin{figure*}[t]
\centering
\begin{tabular}{cc}
\includegraphics[width=2in]{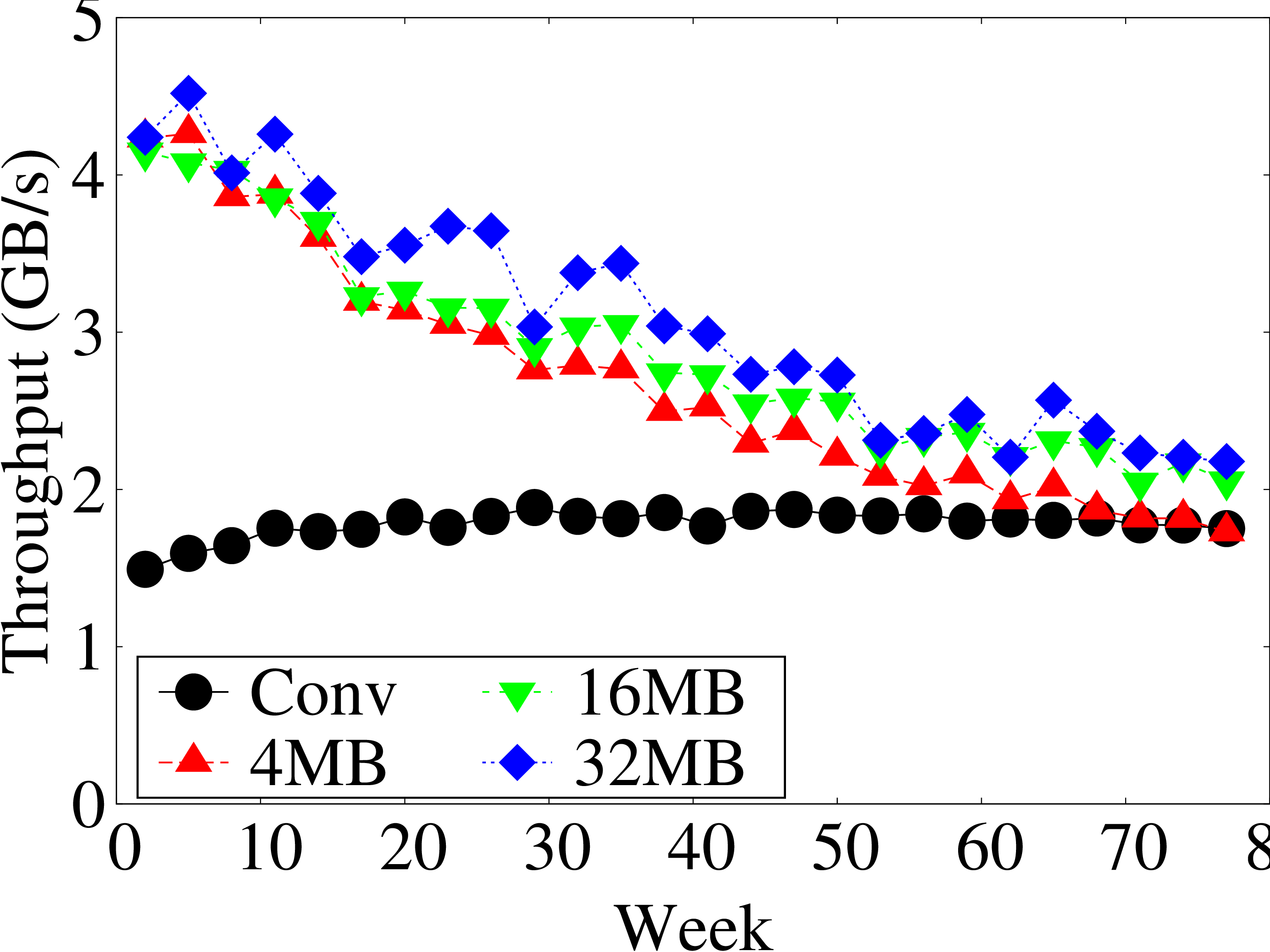} & 
\includegraphics[width=2in]{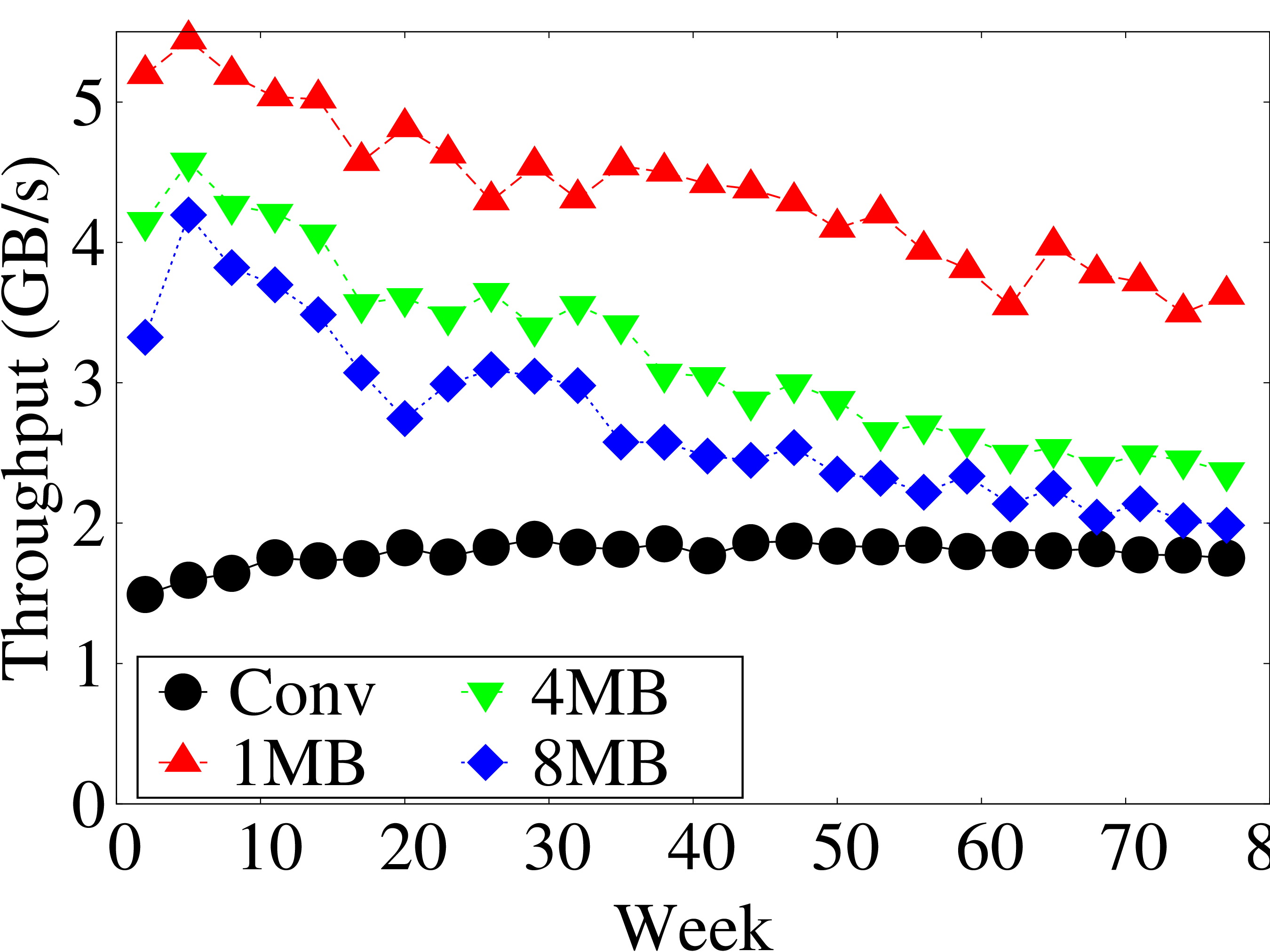} \\
\mbox{\small(a) \textit{SG1}: varied container size} & 
\mbox{\small(b) \textit{SG1}: varied segment size} \\
\includegraphics[width=2in]{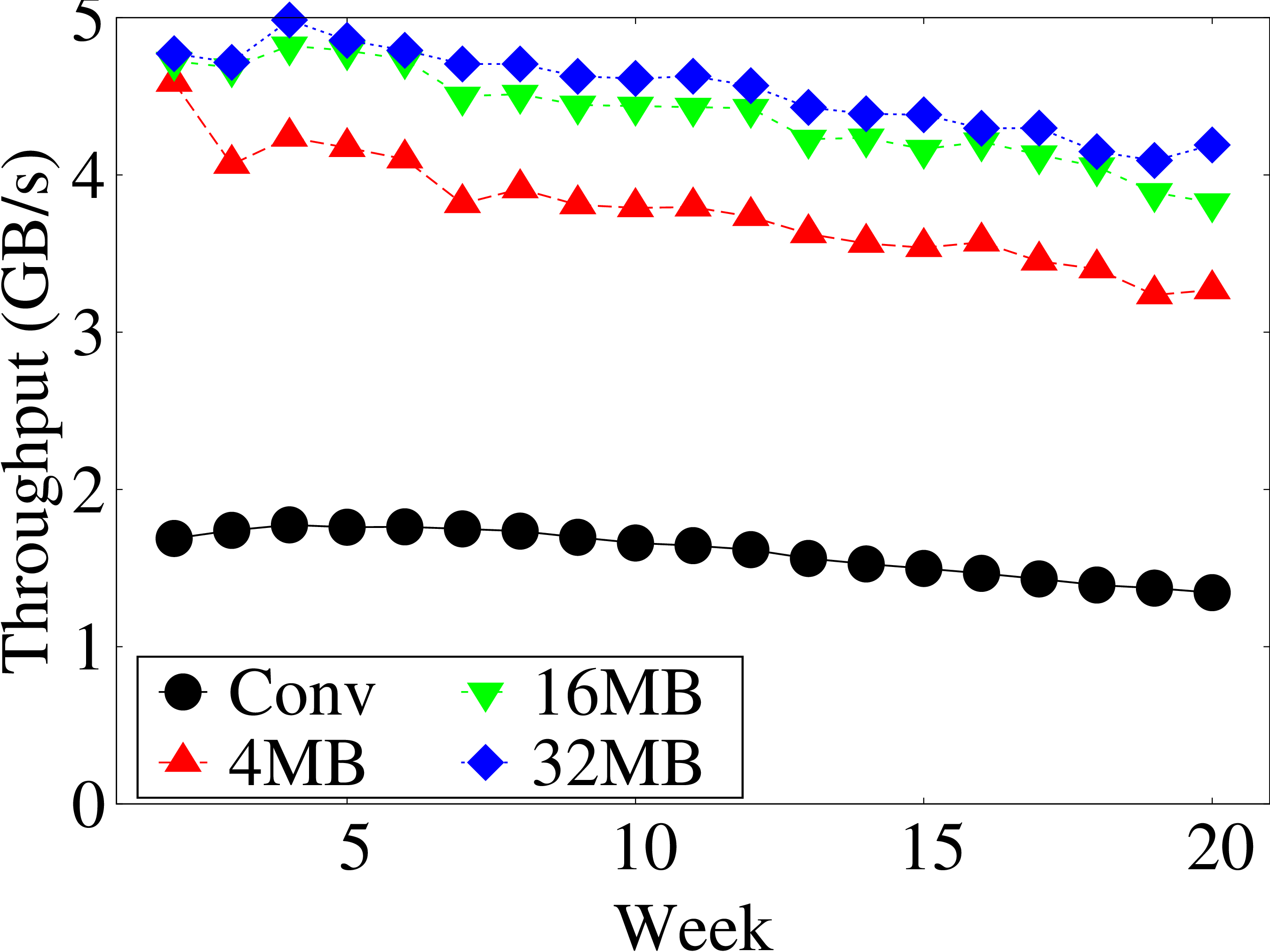} &
\includegraphics[width=2in]{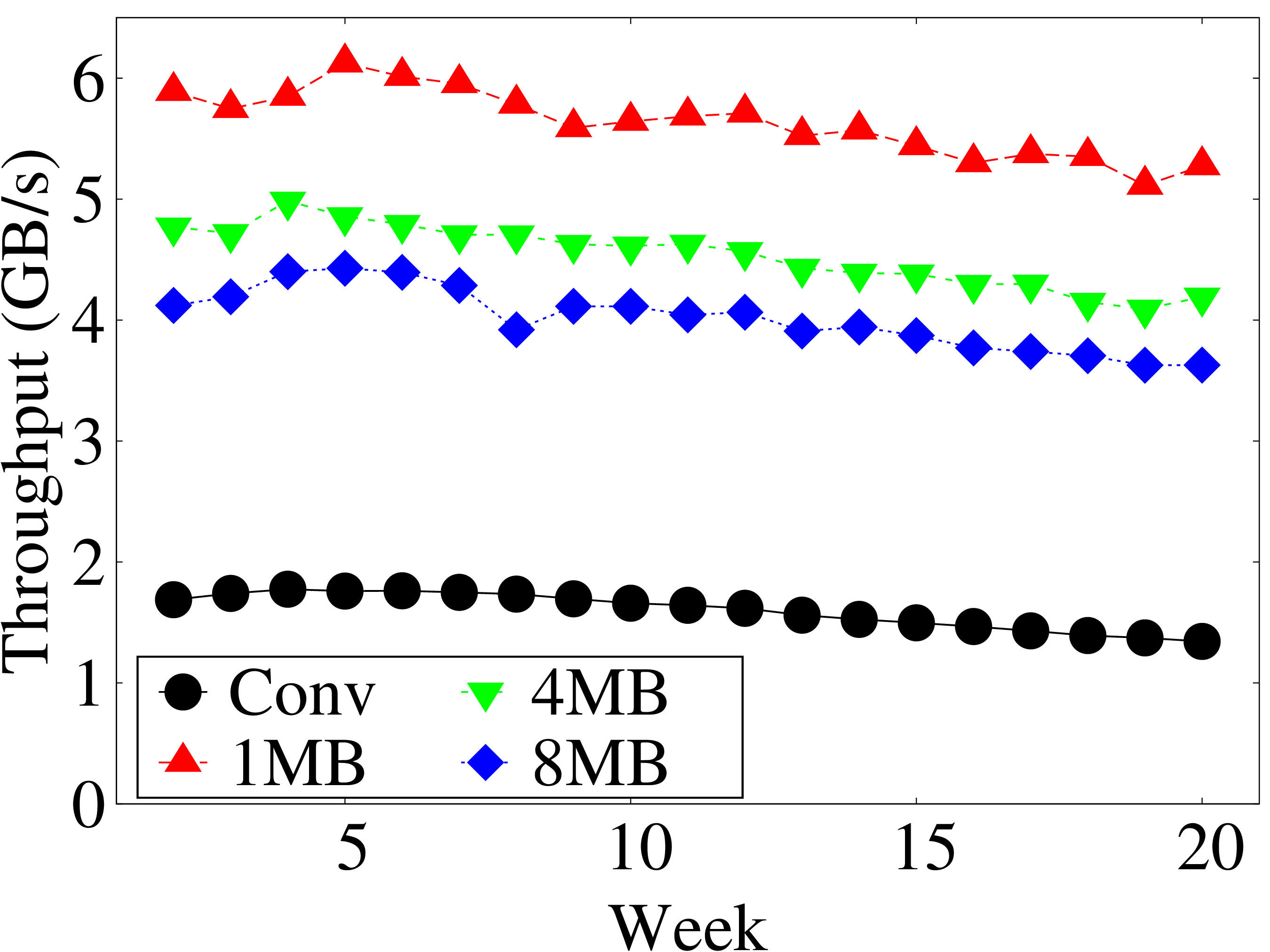} \\
\mbox{\small(c) \textit{GP}: varied container size} &
\mbox{\small(d) \textit{GP}: varied segment size} \\
\includegraphics[width=2in]{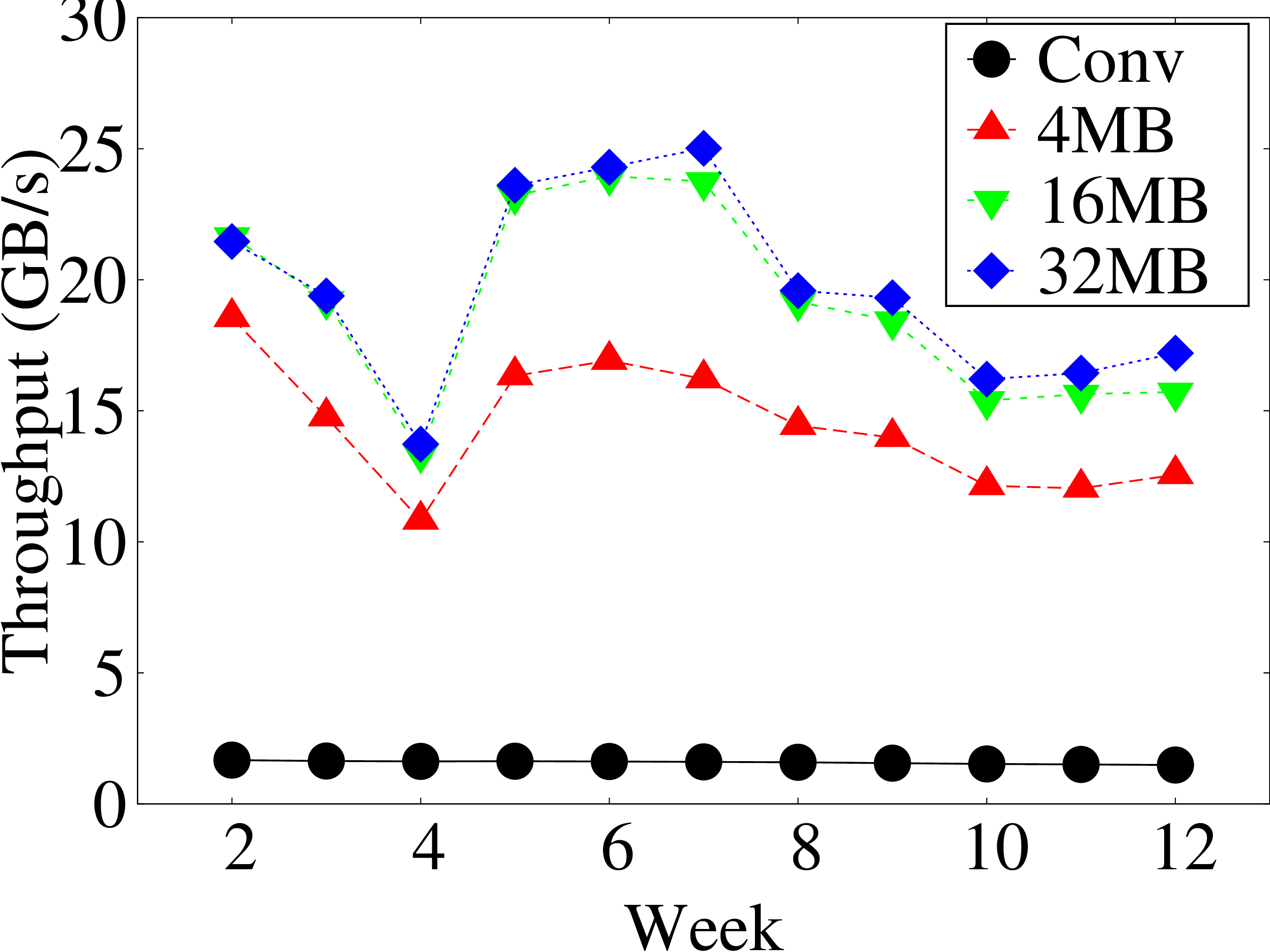} &
\includegraphics[width=2in]{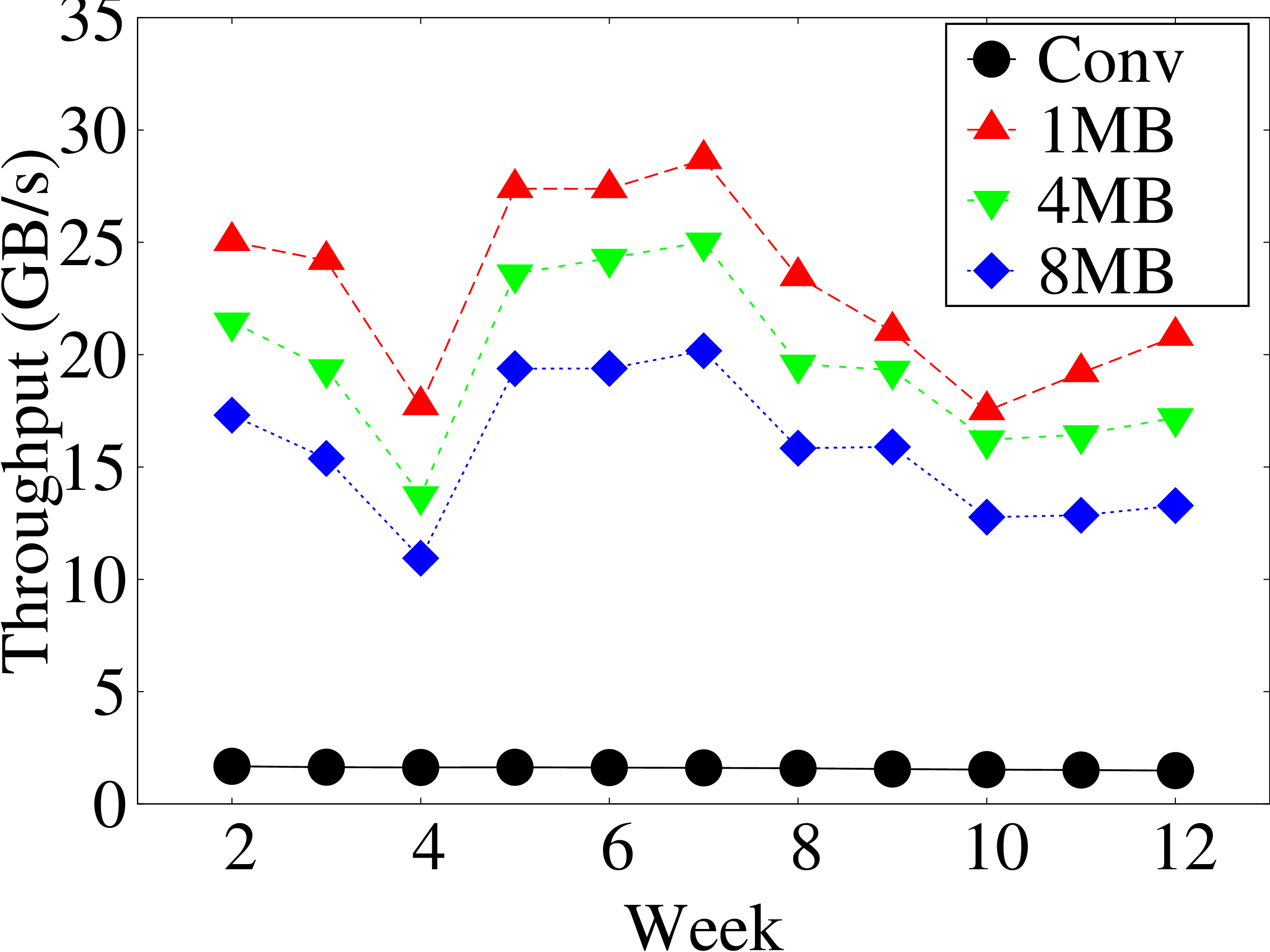} \\
\mbox{\small(e) \textit{VM}: varied container size} &
\mbox{\small(f) \textit{VM}: varied segment size} 
\end{tabular}
\caption{Backup throughput of RevDedup and \textit{Conv} for different
datasets.  We vary the container and segment sizes of RevDedup: figures~(a),
(c), and (e) fix the segment size at 4MB and vary the container size;
figures~(b), (d), and (f) fix the container size at 32MB and vary the segment
size.  The plots start from the second week to take into account inter-version
redundancy.} 
\label{fig:backup_eval}
\end{figure*}



To evaluate the backup throughput of RevDedup, we only measure the backup
performance due to segment-level inline deduplication, since reverse
deduplication is assumed to be done out-of-line.  We will measure the overhead
of reverse deduplication in Section~\ref{subsubsec:reverse_eval}. 

Before each experimental run of a dataset, we format the file system without
any data.  We submit each backup using {\tt inlinededup} (see
Section~\ref{subsec:modules}), and measure the duration starting from when the
in-memory deduplication index is built until all segments are packed into
containers and written to disk.  The backup throughput is calculated as the
ratio of the original backup size to the measured duration.
Figure~\ref{fig:backup_eval} shows the backup throughput results, 
which we elaborate below. 


\paragraph{Performance trends of \textit{Conv} and RevDedup:} At the
beginning, RevDedup has significantly higher backup throughput than the raw
write throughput (e.g., 4$\times$ for \textit{SG1} as shown in
Figure~\ref{fig:backup_eval}(a)).  The throughput decreases over time, as
we make more content changes to the backups and hence introduce more unique
data.  Both synthetic datasets \textit{SG1} and \textit{GP} show similar
trends, due to the ways of how we inject changes to the backups.  

The real-world dataset \textit{VM} shows throughput fluctuations because of
the varying usage patterns.  For example, Week~4 shows a sudden throughput
drop because there was an assignment deadline and students made significant
changes to their VMs.  Despite the fluctuations, RevDedup still achieves much
higher backup throughput than the raw write throughput.  We note that RevDedup
can reach an exceptionally high throughput of up to 30GB/s.  The reason is
that our \textit{VM} dataset contains a large volume of duplicate segments and
null segments, both of which can be eliminated on the write path. 

Although \textit{Conv} has higher reduction of storage space than RevDedup
with only segment-level inline deduplication (see
Figure~\ref{fig:dedup_ratio}), it has much lower backup throughput, and its
throughput is fairly stable.  To explain the results, we measure two
sub-operations of a backup operation: index lookups and data writes (note that
data writes include packing segments to containers and writing containers to
disk).   With multi-threading (see Section~\ref{subsec:improvements}), both
sub-operations are carried out simultaneously, and hence the backup
performance is bottlenecked by the slowest sub-operation.  We consider a
special case when we store the backup of the 2nd week for \textit{SG1}. 
We configure \textit{Conv} with segment size 4KB and RevDedup with varying
segment sizes, while both schemes have container size fixed at 32MB.
Table~\ref{tab:conv_breakdown} provides a time breakdown for the two
sub-operations.  We observe that for \textit{Conv}, although its deduplication
index is kept in memory, its small deduplication units significantly increase
the lookup time and make the index lookups become the bottleneck.  Even though
we inject more unique data to the backups over time, the backup throughput
remains bottlenecked by index lookups.  On the other hand, RevDedup has much
less index lookup overhead with larger segments.  We also note that 
\textit{Conv} has higher data write time than RevDedup, because it adds a much
larger number of small segments into containers. 

We emphasize that as more unique data is added, RevDedup eventually has its
backup throughput dropped below \textit{Conv}.   Nevertheless, the backup
throughput of RevDedup is lower bounded by the baseline for unique data (see
Section~\ref{subsec:baseline}). 

\begin{table}[t]
\centering
\begin{tabular}{|c||c||c|c|c|}
\hline
           &  & \multicolumn{3}{|c|}{RevDedup} \\
\cline{3-5}
           & \textit{Conv} & 1MB & 4MB & 8MB \\
\hline
Index Lookups (sec) & 2.562 & 0.032 & 0.012 & 0.008 \\
\hline
Data Writes (sec) & 0.961 & 0.501 & 0.617 & 0.646 \\
\hline
\end{tabular}
\caption{Time breakdown of writing the second backup of \textit{SG1} for
\textit{Conv} with segment size 4KB and RevDedup with varying segment sizes,
where the container size is fixed at 32MB.}
\label{tab:conv_breakdown}
\end{table}


\paragraph{Effect of container size:} While a larger container size implies
fewer write requests and hence better data write performance, the gains in
backup throughput is insignificant due to the inevitable indexing overhead in
deduplication.  For example, for \textit{SG1} in
Figure~\ref{fig:backup_eval}(a) and \textit{GP} in
Figure~\ref{fig:backup_eval}(c), the backup throughput of RevDedup increases
by only 9\% and 16\% (averaged over all weeks) when the container size
increases from 4MB to 16MB, respectively. 

\paragraph{Effect of segment size:}  A large segment size reduces the
deduplication opportunity, and hence RevDedup writes more data to disk. Since
the data write dominates the backup performance of RevDedup (see
Table~\ref{tab:conv_breakdown}), its backup throughput drops as the segment
size increases.  For example, for \textit{SG1} in
Figure~\ref{fig:backup_eval}(b), the backup throughput drops by 38\%
(averaged over all weeks) when the segment size increases from 1MB to 8MB. 

\subsubsection{Restore}
\label{subsubsec:restore_eval}

\begin{figure*}[t]
\centering
\begin{tabular}{cc}
\includegraphics[width=2in]{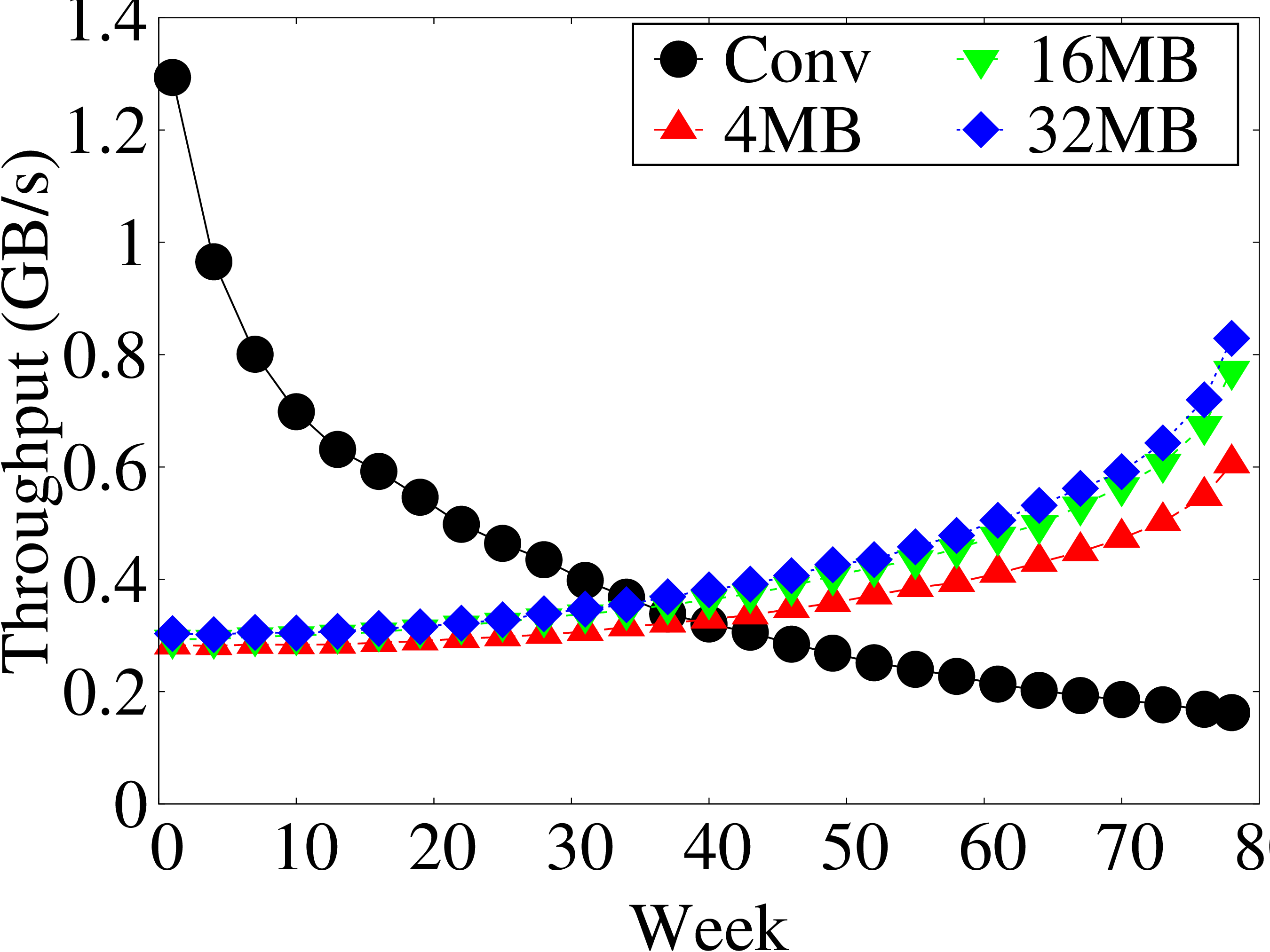} & 
\includegraphics[width=2in]{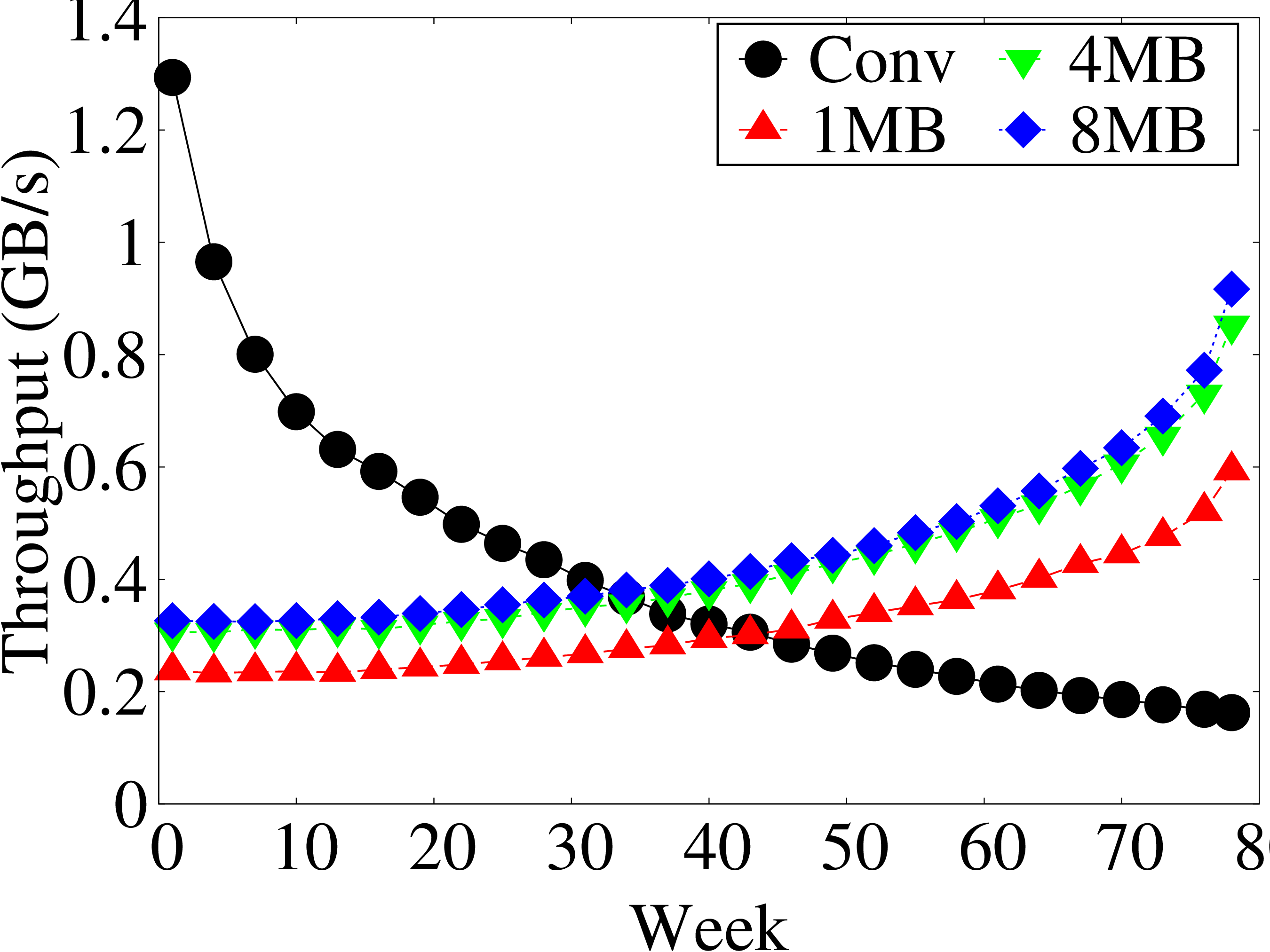} \\
\mbox{\small(a) \textit{SG1}: varied container size} & 
\mbox{\small(b) \textit{SG1}: varied segment size} \\
\includegraphics[width=2in]{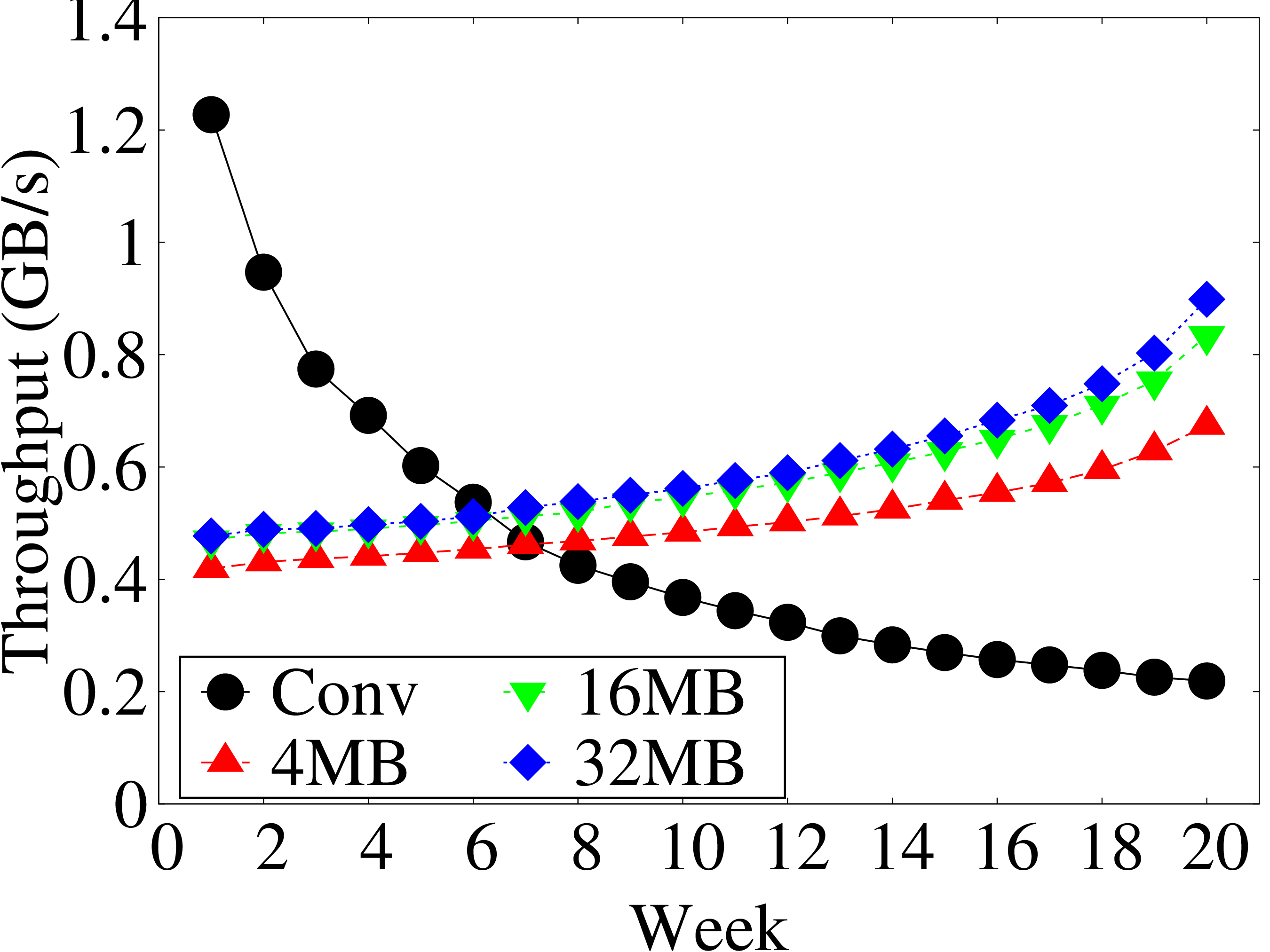} &
\includegraphics[width=2in]{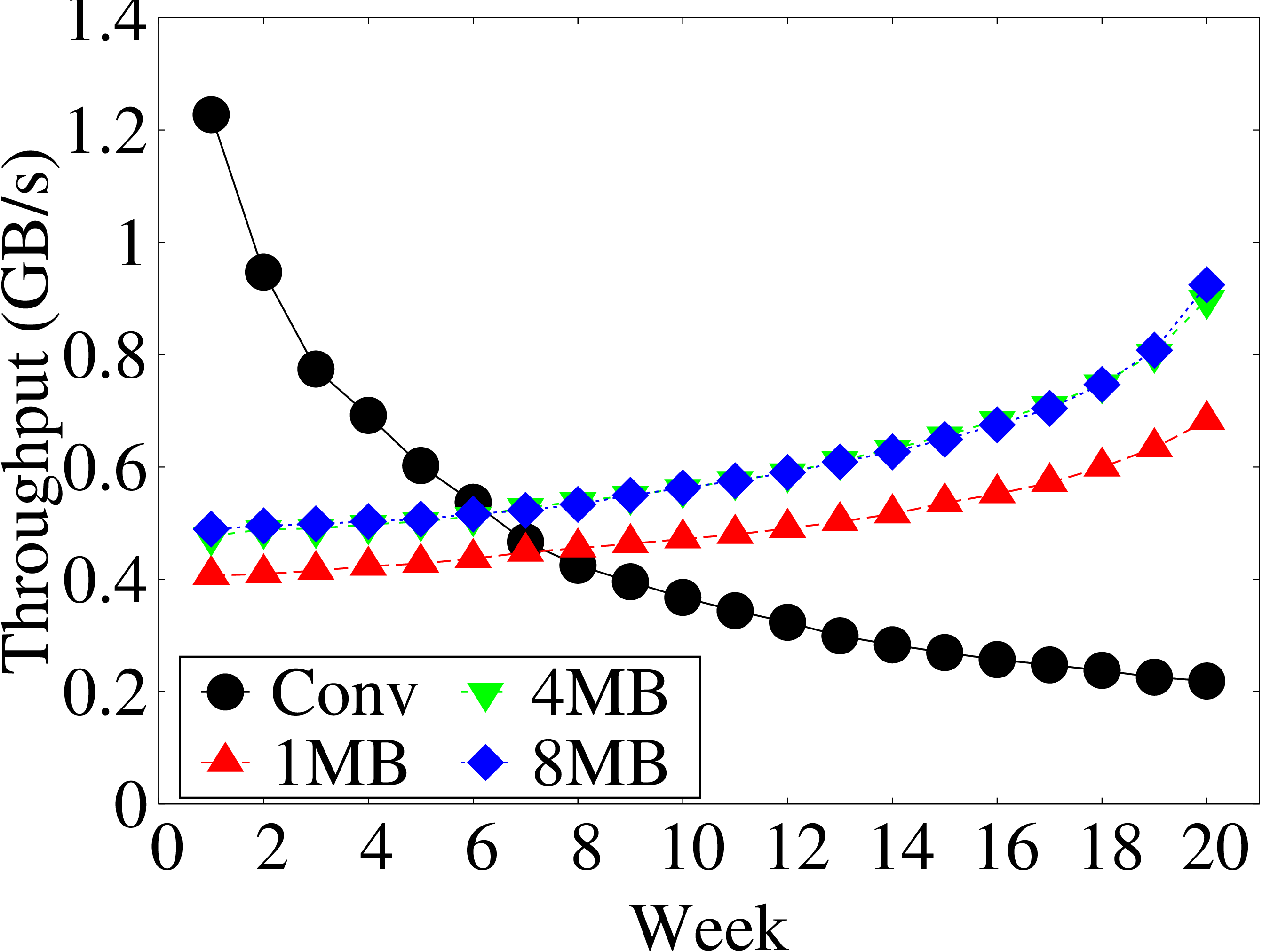} \\
\mbox{\small(c) \textit{GP}: varied container size} &
\mbox{\small(d) \textit{GP}: varied segment size} \\
\includegraphics[width=2in]{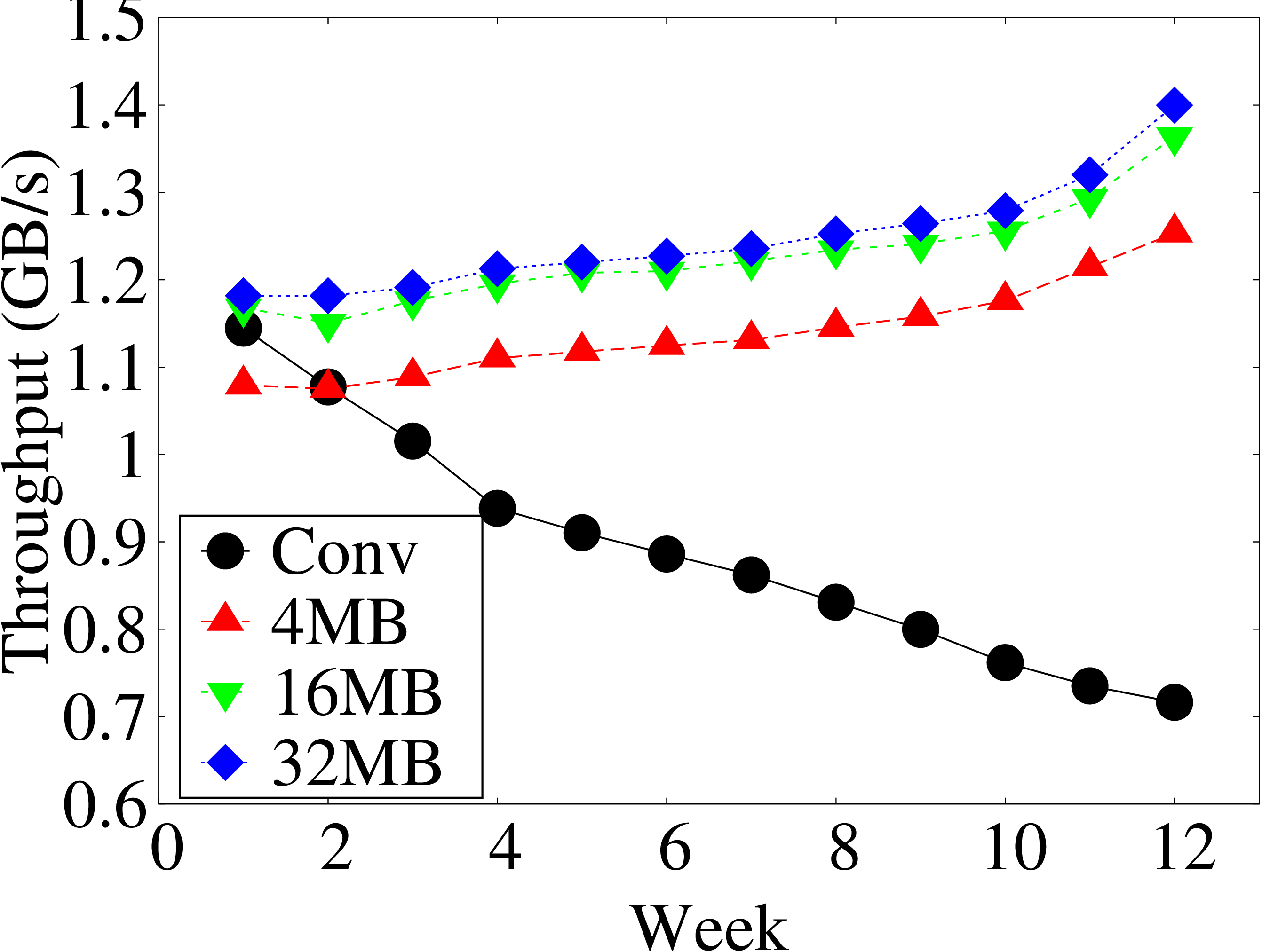} &
\includegraphics[width=2in]{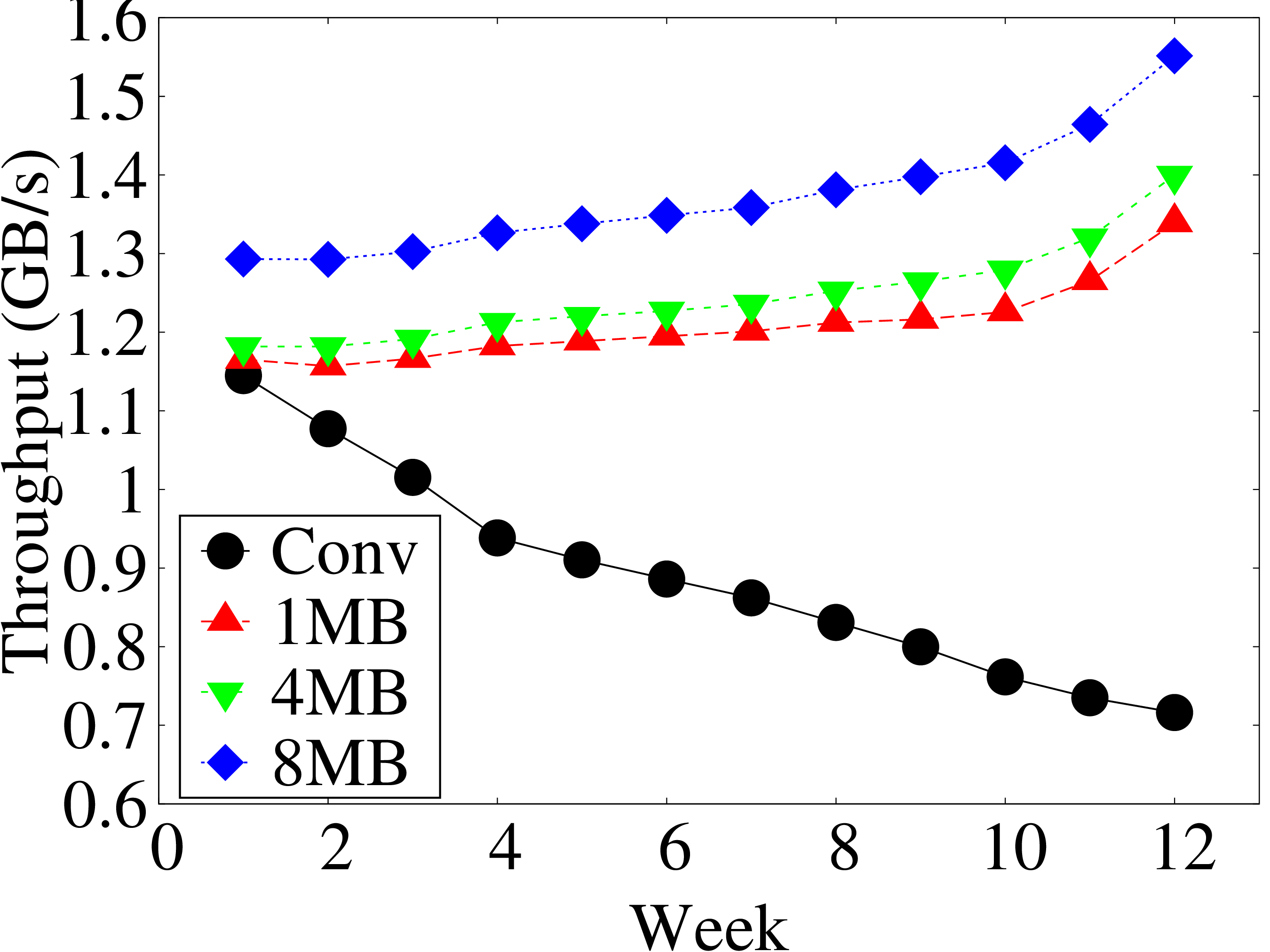} \\
\mbox{\small(e) \textit{VM}: varied container size} &
\mbox{\small(f) \textit{VM}: varied segment size} 
\end{tabular}
\caption{Restore throughput of RevDedup and \textit{Conv}, corresponding to
the settings in Figure~\ref{fig:backup_eval}.}
\label{fig:restore_eval}
\end{figure*}

After writing all backups, we restore each backup using the module 
{\tt restore} (see Section~\ref{subsec:modules}).  We measure the restore
throughput of RevDedup and \textit{Conv} as the ratio of the original backup
size to the restore time.  For RevDedup, we first perform out-of-line reverse
deduplication on the backups, so that the restore performance of both RevDedup
and \textit{Conv} is measured when they have comparable storage efficiency
(see Section~\ref{subsec:storage}).  Figure~\ref{fig:restore_eval} shows the
restore throughput results for various container sizes and segment sizes,
corresponding to the settings of Figure~\ref{fig:backup_eval}.  The results
are elaborated as follows. 

\paragraph{Performance trends of \textit{Conv} and RevDedup:}  
\textit{Conv} suffers from data fragmentation and hence its restore throughput
decreases for more recent backups (e.g., by 86\% from Week~2 to Week~78 in
Figure~\ref{fig:restore_eval}(a)), while RevDedup shifts data fragmentation to
older backups and maintains high restore throughput for the latest backups.
For instance, from Figures~\ref{fig:restore_eval}(a) and
\ref{fig:restore_eval}(c), the restore throughput for the latest backup of
RevDedup is 5$\times$ and 4$\times$ that of \textit{Conv} for \textit{SG1}
and \textit{GP}, respectively.   The throughput values are smaller than the
raw read throughput, mainly due to data fragmentation caused by segment-level
inline deduplication. 


For the real-world dataset \textit{VM}, we see similar trends of restore
throughput for both RevDedup and \textit{Conv}.  However, the restore
throughput of RevDedup goes beyond the raw read throughput (see
Figures~\ref{fig:restore_eval}(e) and \ref{fig:restore_eval}(f)).  The reason
is that the VM images contain a large number of null chunks, which are
generated on the fly by RevDedup rather than read from disk.  We expect that
the restore throughput drops as the number of null chunks decreases.

\paragraph{Effect of container size:}  The restore throughput increases with
the container size as the number of read requests decreases.  For example,
for \textit{SG1} in Figure~\ref{fig:restore_eval}(a) and \textit{GP} in
Figure~\ref{fig:restore_eval}(c), the restore throughput of RevDedup increases
by 11.8\% and 14.4\% (averaged over all weeks) when the container size
increases from 4MB to 16MB, respectively.  However, further increasing the
container size from 16MB to 32MB shows only marginal gains. 

\paragraph{Effect of segment size:}  A larger segment size increases the
restore throughput, as it mitigates data fragmentation.  For example, for
\textit{SG1} in Figure~\ref{fig:restore_eval}(b), the restore throughput for
the latest backup of RevDedup increases by 39.3\% (averaged over all weeks)
when the segment size increases from 1MB to 8MB.  The trade-off of using
larger segments is that it reduces both storage efficiency and backup
throughput. 

\subsubsection{Reverse Deduplication Overhead}
\label{subsubsec:reverse_eval} 

We now evaluate the reverse deduplication throughput, defined as the ratio of
the original backup size to the reverse deduplication time.  Recall that we
set the default live window length as one backup.  After we submit a new
backup, we perform reverse deduplication on its previous backup.  We measure
the time of reading the containers that have non-shared segments of the
previous backup and writing the compacted segments without removed chunks to
disk.  

\begin{figure}[t]
\centering
\includegraphics[width=3in]{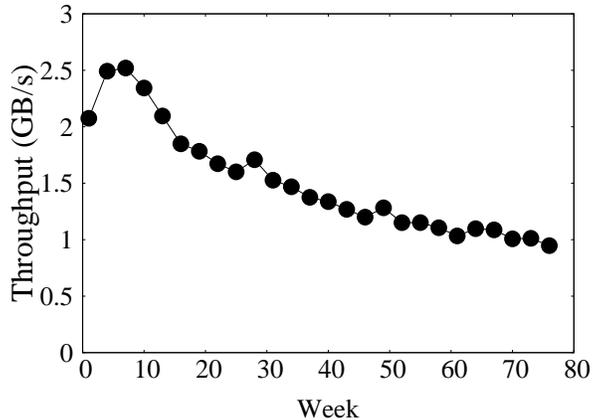}
\caption{Reverse deduplication throughput of RevDedup.}
\label{fig:reverse_eval}
\end{figure}

Figure~\ref{fig:reverse_eval} shows the reverse deduplication throughput for
\textit{SG1} starting from Week~1.  Week~1 has lower throughput than the next
few weeks, due to the following reason. 
Initially, many containers are mixed with shared and non-shared
segments, so we load such containers and separate the shared and non-shared
segments into different containers (see Section~\ref{subsubsec:removal}).  
Later we only load the containers whose
segments change from shared to non-shared, plus the containers that have
non-shared segments associated with the backup on which we apply reverse
deduplication.  We also see that the throughput drops as the amount of unique
data increases, yet it is lower bounded by about half of the baseline
read/write throughput (see Section~\ref{subsec:baseline}) since the whole
backup is read for chunk removal and rewritten to disk (assuming that the
baseline read and write throughputs are about the same). 

\subsubsection{Gains of Prefetching}

\begin{figure}[t]
\centering
\includegraphics[width=3in]{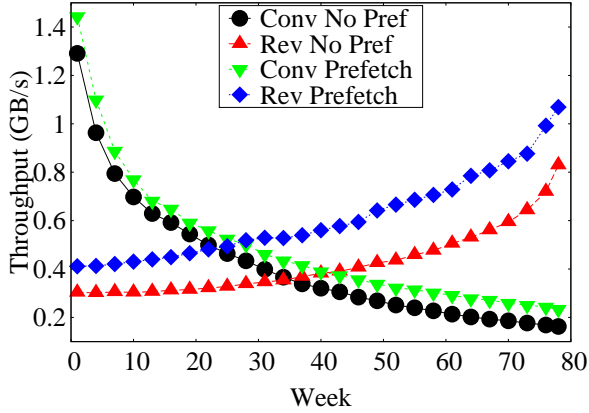}
\caption{Restore throughput of \textit{Conv} and RevDedup with and without
prefetching.}
\label{fig:prefetch}
\end{figure}

We disable prefetching in the previous experiments.  We now evaluate the
restore throughput increases with prefetching enabled.  We focus on
\textit{SG1}. For RevDedup, we fix the segment size at 4MB and the container
size at 32MB.  Figure~\ref{fig:prefetch} shows the results.  We see that
prefetching improves the restore throughput by 23.9\% and 45.8\% (averaged
over all weeks) for \textit{Conv} and RevDedup, respectively.  Note that the
data fragmentation problem, while being mitigated, still manifests in
\textit{Conv}, which still has around 82\% drop in restore throughput from the
first week to the last one.

\subsubsection{Effective Live Window Length}
\label{subsubsec:live}


The live window length defines the number of backups that remain coarsely
deduplicated by segment-level inline deduplication, and determines the
trade-off between storage efficiency and restore performance.  Here, we study
the effect of live window length for the dataset \textit{SG1}.  We first store
all backups of \textit{SG1} and perform reverse deduplication using RevDedup.
We then measure the restore throughput for each backup.  We vary the live
window length to be 1, 5, and 17 backups (i.e., the archival window lengths
are 77, 73, and 61 backups, respectively). 

\begin{figure}[t]
\centering
\includegraphics[width=3in]{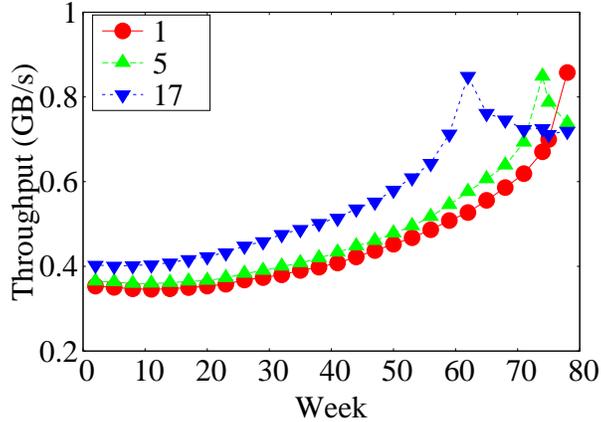} 
\caption{Restore throughput of RevDedup with different live window lengths.}
\label{fig:retention_space}
\end{figure}

Figure~\ref{fig:retention_space} shows the restore throughput results.  The
restore throughput increases over time for backups within the archival window,
since RevDedup shifts data fragmentation to old backups.  On the other hand,
the restore throughput decreases over time for the backups within the live
window (e.g., when the live window is 17 backups), due to data fragmentation
caused by segment-level inline deduplication.  Nevertheless, the drop is
slower than conventional deduplication as the large segment size limits the
overhead of data fragmentation. 

Setting a larger live window increases the restore throughput of backups in
the archival window.  Recall that an archival backup has indirect reference
chains to the oldest live backup.  A larger live window implies that an
archival backup has shorter indirect reference chains.  This reduces the
tracing time when restoring the archival backups. 

The trade-off is that a larger live window length increases the storage space.
For example, the percentage reductions of space saved due to deduplication
drop from 96.6\% to 93.2\% (averaged over all weeks) when the live window
length increases from 1 to 17 backups. 

\subsection{Deletion Overhead}

\begin{figure}[t]
\centering
\begin{tabular}{cc}
\includegraphics[width=2in]{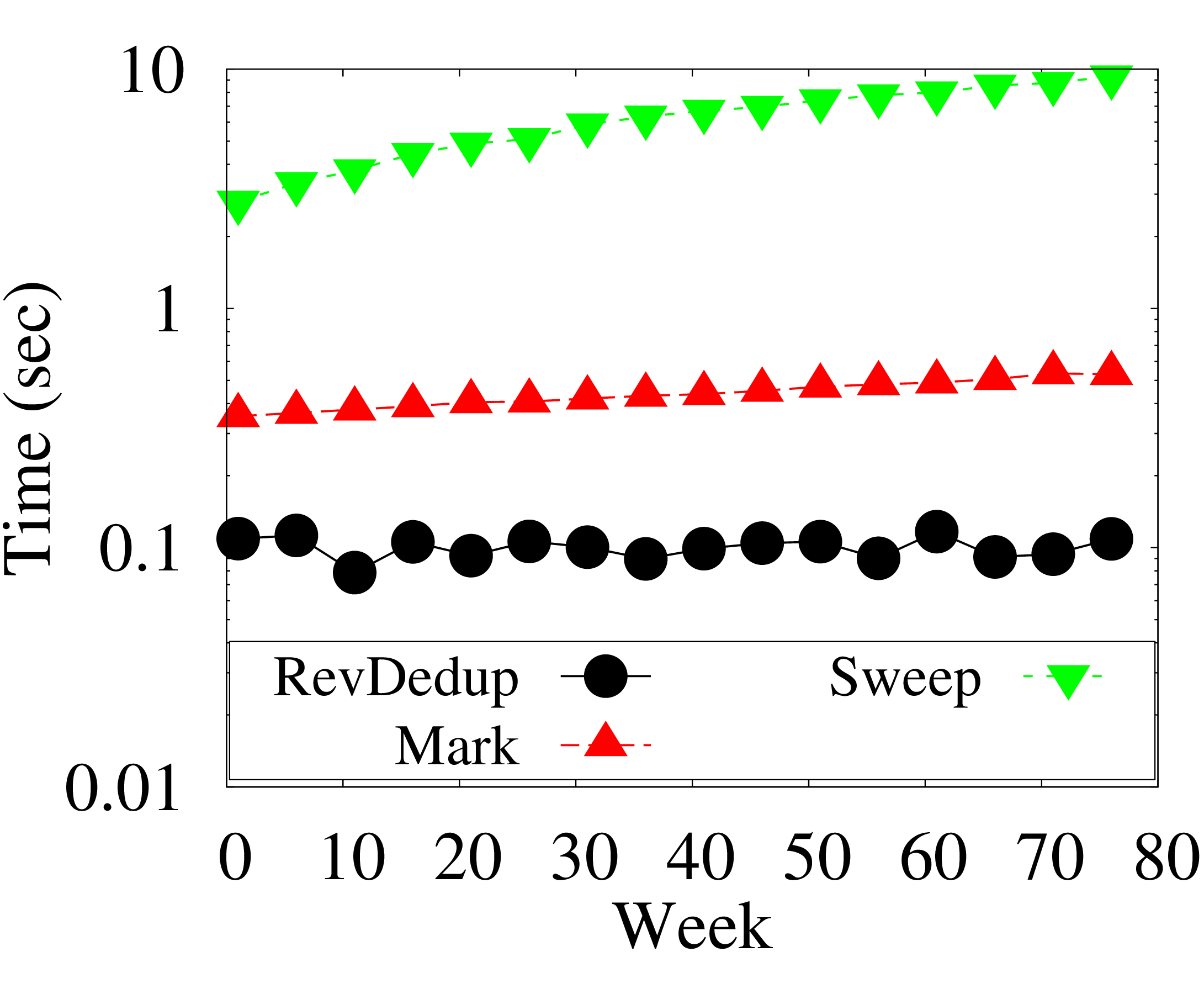} & 
\includegraphics[width=2in]{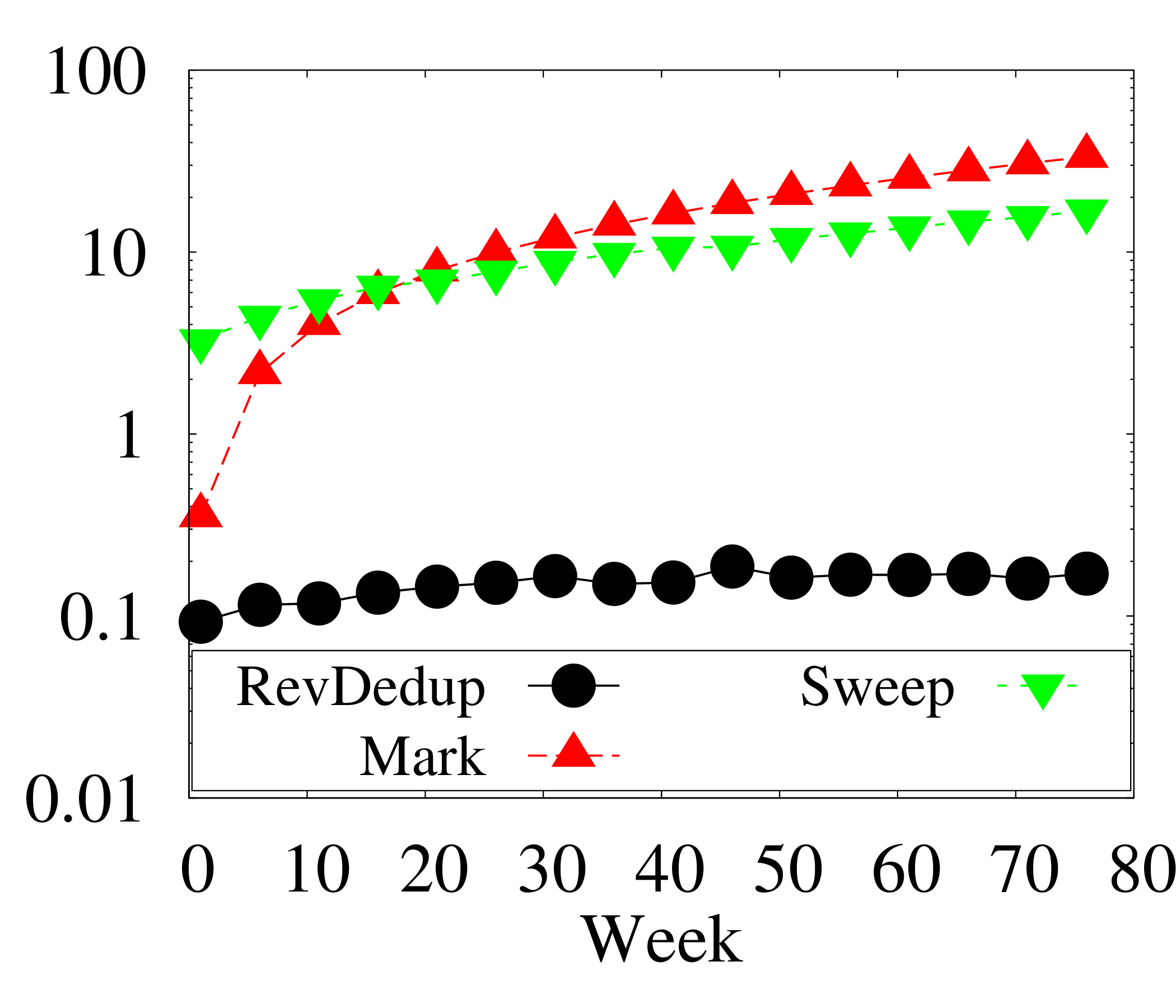} \\
\mbox{\small(a) Incremental Deletion} & 
\mbox{\small(b) Batch Deletion}
\end{tabular}
\caption{Deletion times of RevDedup and mark-and-sweep.}
\label{fig:del}
\end{figure}

We evaluate the deletion overhead of RevDedup and compare it with the
traditional mark-and-sweep approach.  We consider two types of deletion
operations: incremental deletion of the earliest backup and batch deletion of
multiple expired backups.  We first store 78 weeks of backups and perform
reverse deduplication using RevDedup, and then run each type of deletion.
Figure~\ref{fig:del} shows the average results over five runs for the dataset
\textit{SG1}.  We elaborate the results below. 

\paragraph{Incremental deletion:} In this test, we keep deleting a backup from
the series one by one, starting from the earliest backup.  RevDedup simply
deletes the metadata of the deleted backup and the containers whose timestamps
are equal to the creation time of the deleted backup.  On the other hand, 
in the mark-and-sweep approach, the mark phase loads the metadata of the
backup and decrements the reference count of each associated segment, and the
sweep phase scans through all containers to delete the non-referenced segments
and reconstruct the containers with the remaining segments that are not
deleted.  Figure~\ref{fig:del}(a) shows the time breakdown.  The mark phase
incurs small running time as it only processes metadata, while the sweep
phase incurs significant running time as it needs to read and reconstruct the
containers.   RevDedup has significantly smaller deletion time than each
of the mark and sweep phases. 

\paragraph{Batch deletion:} In this test, we delete the $n$ earliest backups,
with $n$ ranging from 1 (only the earliest one) to 77 (all except the most
recent one).  To measure the time of deleting $n$ backups, we first take a
snapshot of the storage partition and store the snapshot elsewhere, perform
the deletion and record the time, and finally restore the snapshot to prepare
for the next deletion.  The deletion processes of both RevDedup and
mark-and-sweep are similar to those in individual deletion.
Figure~\ref{fig:del}(b) shows the time breakdown.  The running time of the
mark phase increases with $n$ since it reads the metadata of more backups,
while the sweep phase has similar running time as in incremental deletion as
it scans through all containers once only.  The deletion time of RevDedup
remains very small. 

\paragraph{Summary:} The two tests show that RevDedup incurs small overhead in
both incremental and batch deletion operations.  The deletion overhead is
amortized over the chunk removal process during reverse deduplication. 
The small batch deletion overhead of RevDedup provides flexibility for
administrators to defer the deletion of expired backups as long as the storage
space remains sufficient. 


\section{Related Work}
\label{sec:related}


\paragraph{Backup:}  Most existing deduplication studies for
backup storage focus on achieving high backup performance.  Deduplication is
first proposed in Venti \cite{quinlan02} for data backup in
content-addressable storage systems.  DDFS \cite{zhu08} and Foundation
\cite{rhea08} maintain fingerprints in a Bloom filter \cite{bloom70} to
minimize memory usage for fingerprint indexing.  DDFS further exploits spatial
locality to cache the fingerprints of blocks that are likely written later.
Other studies \cite{lillibridge09,bhagwat09,kruus10,guo11,xia11,meister13}
exploit workload characteristics to further improve backup performance while
limiting the memory overhead for indexing.  ChunkStash \cite{debnath10} and
Dedupv1 \cite{meister10} store fingerprints in solid state drives to achieve
high-speed fingerprint lookup.  All above approaches build on inline
deduplication, while RevDedup uses out-of-line deduplication to address both
restore and deletion performance.  In particular, Bimodal \cite{kruus10}
uses a hybrid of large and small chunk sizes.  Although seemingly similar to
RevDedup, it dynamically switches between the chunk sizes in inline
deduplication, while RevDedup uses out-of-line deduplication on small-size
chunks.   

\paragraph{Restore:} To mitigate chunk fragmentation in inline
deduplication and hence improve restore performance, Kaczmarczyk {\em et al.}
\cite{kaczmarczyk12} propose context-based rewriting, which selectively
rewrites a small percentage of data for the latest backups.  Nam {\em et al.}
\cite{nam12} measure the fragmentation impact given the input workload and
activate selective deduplication on demand.  Lillibridge {\em et al.}
\cite{lillibridge13} use container capping to limit the region of chunk
scattering, and propose the forward assembly area (similar to caching) to
improve restore performance.  Note that the studies
\cite{kaczmarczyk12,nam12,lillibridge13} only conduct simulation-based
evaluations, while we implement a prototype to experiment the actual I/O
throughput.  SAR \cite{mao14} leverages SSDs to store the chunks referenced by
many duplicate chunks and absorb the random reads to harddisks.  In contrast,
RevDedup does not rely on the use of SSDs. 

The above approaches are designed for backup storage, while iDedup
\cite{srinivasan12} is designed for primary storage and it limits disk seeks
by applying deduplication to chains of continuous duplicate chunks rather than
individual chunks.  

\paragraph{Reclamation:}  Several approaches focus on reducing the reclamation
overhead in inline deduplication systems.  
Guo {\em et al.} \cite{guo11} propose a grouped mark-and-sweep approach that
associates files into different backup groups and limits the scanning to only
a subset of backup groups.  
Botelho {\em et al.} \cite{botelho13} propose a memory-efficient data
structure for indexing chunk references.  Strzelczak {\em et al.}
\cite{strzelczak13} extend HYDRAstor \cite{dubnicki09} with concurrent
deletion to minimize the interference of background sweeping to ongoing
writes.  Simha {\em et al.} \cite{simha13} limit the number of reclaimed
chunks and ensure that the reclamation overhead to be proportional to the size
of incremental backups.  

\section{Conclusions} 
\label{sec:conclusion}

We explore the problem of achieving high performance in essential operations
of deduplication backup storage systems, including backup, restore, and
deletion, while maintaining high storage efficiency.  We present RevDedup, an
efficient hybrid inline and out-of-line deduplication system for backup
storage.  The key design component of RevDedup is reverse deduplication, which
removes duplicates of old backups out-of-line and mitigates fragmentation of
latest backups.  We propose heuristics to make reverse deduplication
efficient: (1) limiting the deduplication operation to consecutive backup
versions of the same series, (2) using two-level reference management to keep
track of how segments and chunks are shared, and (3) checking only non-shared
segments for chunk removal.  We extensively evaluate our RevDedup prototype
using synthetic and real-world workloads and validate our design goals.  
We plan to release the source code of our RevDedup prototype in the final
version of the paper.

\begin{footnotesize}
\bibliographystyle{plain}
\bibliography{paper}
\end{footnotesize}

\end{document}